\documentclass{aa}  
\usepackage{natbib} 
\bibpunct{(}{)}{;}{a}{}{,} 
\usepackage{graphicx}
\usepackage{txfonts}
\usepackage{lipsum}
\usepackage{subcaption}     
\usepackage{lscape}   
\usepackage{placeins} 
\usepackage{hyperref}
\hypersetup{
    colorlinks=true,
    linkcolor=blue,
    filecolor=magenta,      
    urlcolor=blue,
    citecolor=blue,
    pdfpagemode=FullScreen,
    }

\begin{document}

   \title{Toward Bayesian photometric cosmic chronometers: Application to VIPERS}
   \titlerunning{Cosmic chronometers with VIPERS}

   \author{S. Pradhan\inst{1,2,3}\corrauth{spradhan@ice.csic.es}
        \and P. Renard\inst{1,2,4}\email{renard@ieec.cat}
        \and E. Gazta\~naga\inst{1,2,3}\email{enrique.gaztanaga@port.ac.uk}
        \and M. Siudek \inst{1, 5}\email{msiudek@iac.es}
        \and T. Moutard \inst{6}\email{Thibaud.Moutard@esa.int}
        }

   \institute{Institute of Space Sciences (ICE, CSIC), Campus UAB, Carrer de Can Magrans, s/n, 08193 Barcelona, Spain
   \and Institut d\'~Estudis Espacials de Catalunya (IEEC), 08034 Barcelona, Spain
   \and Institute of Cosmology and Gravitation (ICG), University of Portsmouth, Portsmouth, PO1 3FX, UK
   \and Department of Astronomy, Tsinghua University, Beijing 100084, China
   \and Instituto de Astrof\'{\i}sica de Canarias (IAC), Departamento de Astrof\'{\i}sica, Universidad de La Laguna (ULL), 38200 La Laguna, Tenerife, Spain
   \and European Space Agency (ESA), European Space Astronomy Centre (ESAC), Camino Bajo del Castillo s/n, 28692 Villaneuva de la Ca\~nada, Madrid, Spain}
   
   \date{Received June 5 2026}
 
  \abstract
   {The cosmic chronometer (CC) method provides a direct measurement of the expansion history of the universe, $H(z)$, from the differential ages of passively evolving galaxies. 
   Unlike standard geometrical probes, which constrain integrals over the expansion history, CCs directly probe $dz/dt$ and therefore provide a complementary and largely model-independent measurement of the cosmic expansion rate. 
   However, most CC analyses rely on high-quality spectroscopy to select passive galaxies, measure age-sensitive spectral features, and control stellar-population systematics.}
   {We build on existing works that use the D4000 spectral break as a proxy for measuring galaxy ages and apply it to a photometry-selected galaxy sample from VIPERS PDR2 in the range $0.5 \le z \le 0.8$. Our goal is to extend the scope of the standard CC framework to photometric surveys.} 
   {To achieve this, we first selected a massive and passive galaxy sample using rest-frame colors and mass. Second, we designed a Bayesian framework to infer full galaxy age posteriors in fine redshift bins, using a D4000-age-metallicity grid from stellar population synthesis (SPS) models. We marginalized over metallicity, using a Gaussian metallicity prior ($\log(Z/Z_\odot) = 0.13 \pm 0.032$) to break the D4000-age-metallicity degeneracy. Subsequently, we derived age-difference posteriors between the $i^{\rm th}$ and $(i+n)^{\rm th}$ redshift bins by convolving their age posteriors to accurately propagate non-Gaussian features. Finally, using the median and errors extracted from the differential age posteriors, we calculated the inverse-variance-weighted average $H(z)$ over our selected redshift range.}
   {We obtain $H(z=0.65)=93.68\pm28.27\,{\rm (stat.)}\pm10.67\,{\rm (syst.)}\ {\rm km\,s^{-1}\,Mpc^{-1}}$, which is consistent with existing spectroscopic CC measurements and with the standard $\Lambda$ cold dark matter ($\Lambda$CDM) model prediction from Planck at the same redshift.}
   {This result provides a proof of concept for extending direct $H(z)$ measurements from cosmic chronometers to photometric and spectro-photometric surveys, where larger samples can compensate for lower spectral resolution, provided that passive-galaxy selection, metallicity priors, and stellar-population systematics are carefully controlled.}


   \maketitle

\nolinenumbers

\section{Introduction}
\label{S:1}

The expansion history of the Universe, $H(z)$, is one of the central observables of modern cosmology.  
It encodes the relative contributions of matter, radiation, curvature, and any component or physical mechanism responsible for the late-time acceleration of cosmic expansion.  
Accurate measurements of $H(z)$ are therefore essential not only for constraining the parameters of the standard $\Lambda$ cold dark matter ($\Lambda$CDM) model, but also for testing whether cosmic acceleration reflects a cosmological constant, evolving dark energy, modified gravity, or a more global effect associated with the structure and boundary conditions of spacetime.

In the early 2000s, the standard cosmological framework was already shaped by multiple independent probes. Measurements of cosmic microwave background (CMB) anisotropies and the large-scale structure, together with galaxy cluster abundances, yielded precise constraints on the matter density parameter, $\Omega_m$ , in a low-density and nearly flat universe \citep[e.g.,][]{Peacock_2001}.  
At the same time, the Hubble Space Telescope (HST) Key Project converged on a value of $H_0 \simeq 72 \pm 8$ km s$^{-1}$ Mpc$^{-1}$, substantially reducing the uncertainty in the Hubble constant \citep{Freedman_2001}.  
Soon after, observations of type Ia supernovae (SNeIa) revealed that the expansion of the Universe is accelerating, implying the presence of dark energy with an equation-of-state parameter close to $\omega=-1$ \citep{Riess_1998, Perlmutter_1999}.

However, most geometrical observables, such as luminosity and angular-diameter distances, are integrals over the inverse Hubble parameter.  
Consequently, rather different expansion histories, $H(z)$, or dark-energy equations of state, $\omega(z)$, may produce nearly degenerate distance--redshift relations \citep{Maor_2001, Jimenez_Loeb_2002}.  
This limitation has motivated the development of methods capable of probing the expansion rate more directly.

One important example is the measurement of the radial baryon acoustic oscillation (BAO) scale, which constrains $H(z)$ through the line-of-sight clustering signal \citep{Gaztanaga_2009}.  
Another is the redshift-drift or Sandage--Loeb test \citep{Sandage_1962, Loeb_1998}, which aims to detect the temporal evolution of cosmological redshifts directly.  
Among these probes, the cosmic chronometer (CC) method is unique in attempting to measure cosmic expansion itself through the differential aging of galaxies.

The use of relative galaxy ages as a direct probe of $dz/dt$, and hence of $H(z)$, was proposed by \cite{Jimenez_Loeb_2002}.  
The method is based on selecting an ensemble of old, massive, passively evolving galaxies at nearby redshifts $z$ and $z+\Delta z$, and estimating their differential age, $\Delta t$.  
Under the assumption of a Friedmann--Lema\^itre--Robertson--Walker (FLRW) metric, the expansion rate can then be written as \citep{Jimenez_Loeb_2002}
\begin{equation}
    H(z) = -\frac{1}{1+z}\frac{dz}{dt}.
    \label{eq:cc}
\end{equation}

Equation~\eqref{eq:cc} is commonly known as the CC equation. Unlike standard distance probes, the CC method does not require us to assume a specific cosmological model for $H(z)$.  
A direct measurement of $H(z)$ is therefore especially valuable because it probes the expansion rate itself rather than an accumulated distance \citep[see,][for a detailed explanation]{Vagnozzi_2021}.
In this sense, cosmic chronometers attempt to measure cosmic expansion directly by tracking the differential aging of galaxies.

This model independence makes CC measurements particularly relevant in the current era of precision cosmology.  
Observations of the CMB, large-scale structure, baryon acoustic oscillations, and type Ia supernovae have established a remarkably successful standard cosmological model, but they have also revealed persistent tensions and degeneracies. 
The most prominent is the so-called Hubble tension: the discrepancy between early Universe inferences of $H_0$ from CMB data and local distance-ladder measurements based on Cepheids and SNeIa \citep{Verde_2019_hubble_tension, DiValentino_2021_hubble_tension}.  
Although unresolved systematics remain possible, the tension has motivated a broad range of extensions to $\Lambda$CDM \citep{Riess_2020_Hubble_tension, Vagnozzi_2022_hubble_tension_alternatives, Perivolaropoulos_2022_hubble_tension_alternatives, Aloni_2022_hubble_tension_alternative}.  
Independent measurements of $H(z)$ at intermediate redshifts are therefore essential for distinguishing between early-time, late-time, and astrophysical explanations. Moreover, CC measurements might be particularly valuable for arbitrating the Hubble tension, a point emphasized in dedicated reviews and analyses of the tension \citep{Vagnozzi_2023, Pedrotti_2026}.

The reliability of the CC method depends critically on the selection of suitable chronometers.  
The ideal galaxies are massive, old, and passively evolving systems with negligible recent star formation, mergers, or gas accretion.  
It has been known since the early works of \cite{Hubble_1936} that galaxy morphologies correlate with their colors and with other derived properties of their stellar population \citep{Mignoli_2009}. And since then, alongside morphological cuts, multiple other methods have been proposed in the literature to separate star-forming and passively evolving galaxies. One of the most popular methods is color-color diagrams introduced by \cite{Williams_2009} with UVJ colors. Based on this method, \cite{Ilbert_2013_proc} used NUVrJ diagrams, and \cite{Arnouts_2013} used NUVrK diagrams to separate red or passively evolving galaxies from blue or star-forming ones. In a similar way, \cite{Peng_2010} introduced color-mass diagrams and \cite{Pozzetti_2010} introduced a specific star formation rate ($\rm sSFR$) criterion. Following \cite{Ilbert_2010} and \cite{Pozzetti_2010}, \cite{Moresco_2013} used $\log($\rm sSFR$) < -2 \,(\mathrm{Gyr}^{-1})$ to select ``quiescent'' galaxies. They also note how \cite{Ilbert_2010} show this cut directly corresponds to a $NUV - r$ cut of 3.5. All these selection methods do not fully overlap \citep{Renzini_2006_Passive}, and selections based on a single method are not stringent enough to remove star-forming outliers, which typically introduce 10 to 30\% contamination \citep{Borghi_2022a}. In this work, we therefore use a combination of colors and spectral energy distribution (SED)-fitting data (e.g., stellar mass, sSFR) to select a sufficiently passive galaxy sample.

For such populations, the evolution of their spectra is driven mainly by stellar aging, making stellar population synthesis (SPS) models sufficiently predictive for differential-age measurements.  
This is a less demanding task than determining absolute galaxy ages, because many systematic uncertainties partly cancel when age differences are taken \citep{Moresco_2011, Moresco_2012b, Borghi_2022a}.
Nevertheless, residual systematics from SPS models, stellar libraries, metallicity, and star-formation histories must still be propagated carefully into the final $H(z)$ uncertainties \citep{Moresco_2020}.

Early CC studies estimated galaxy ages by fitting SPS models directly to spectra \citep{Jimenez_2003, Simon_2005, Stern_2010}.  
A major development was the use of the 4000\,\AA\ spectral break, D4000, as an age-sensitive observable for passive galaxies \citep{Moresco_2011}.  
D4000 is relatively easy to measure, has a high signal-to-noise ratio compared with many individual absorption features, and is less sensitive to flux calibration, dust reddening, and sky residuals than full spectral fitting.  
In the passive regime relevant for CC analyses, the D4000-age relation is approximately linear, although its slope depends on metallicity, star-formation history, stellar libraries, and the adopted SPS model.

Most D4000-based CC analyses therefore proceed by calibrating the D4000-age slope, $A({\rm SFH},Z)$, for different metallicities and SPS assumptions, and then propagating model uncertainties into $H(z)$ \citep{Moresco_2011, Moresco_2012a, Moresco_2012b, Moresco_2013, Moresco_2015, Moresco_2016, Moresco_2018, Loubser_2025b}.  
This approach has been successful, but it typically relies on high-quality spectroscopy to both select passive galaxies and constrain metallicity.  
Alternative methods, including full spectral fitting and Lick or pseudo-Lick indices, can provide more detailed stellar-population information, but require high signal-to-noise spectra and careful calibration \citep{Zhang_2014_fsft, Ratsimbazafy_2017, Borghi_2022a, Borghi_2022b, Jiao_2023}.

The need for high-quality spectroscopy limits the statistical power and survey volume of CC studies.  
It also motivates the question addressed in this work: can the CC framework be extended to photometric or spectro-photometric surveys, where much larger samples are available but the spectral information is weaker?  
Recent work by \cite{Jimenez_2023} shows that photometric data, calibrated with spectroscopic training sets, can be used to infer age-redshift relations for passive galaxies.  
Here, we pursue a complementary approach: we retain the physical transparency of the D4000 method but recast the age inference in a Bayesian framework that explicitly propagates non-Gaussian uncertainties and metallicity degeneracies. 

We test this framework using the VIMOS Public Extragalactic Redshift Survey (VIPERS) Public Data Release 2 \citep[PDR2;][]{VIPERS_PDR2_2018}, which provides spectroscopy and multiband photometry over the redshift range of interest.  
VIPERS is particularly suitable for this proof of concept because it contains a large intermediate-redshift galaxy sample, has well-studied passive-galaxy selections \citep{Moutard_2016b, Siudek2016}, and covers the D4000 feature over $0.5\le z\le0.8$.  
Although its spectral resolution, $R\simeq220$, is lower than that normally used in precision CC studies, this makes it a useful test case for methods intended for photometric-quality data.

This paper is organized as follows.  
Section~\ref{S:2} describes the VIPERS data.  
Section~\ref{S:3} presents the selection of massive, passively evolving galaxies.  
Section~\ref{S:4} describes the SPS models and the D4000--age--metallicity grids.  
Section~\ref{S:5} introduces the Bayesian age-inference framework and the construction of age-difference posteriors.  
Section~\ref{S:6} presents the resulting $H(z)$ measurement and error budget.  
Section~\ref{S:7} discusses the implications, limitations, and applicability to future photometric surveys.  
We conclude in Section~\ref{S:8}.

\section{Data}
\label{S:2}

The VIMOS Public Extragalactic Redshift Survey is a large spectroscopic galaxy survey designed to map the large-scale structure and galaxy population of the Universe at intermediate redshift, with a median redshift of $z \simeq 0.7$, and it is a deep, highly dense spectroscopic survey \citep{VIPERS_PDR1_2014, VIPERS_2014_Guzzo}. VIPERS targets the general galaxy population at $0.5 \lesssim z \lesssim 1.2$ over two CFHTLS-Wide fields (W1 and W4), combining a relatively bright flux limit with a simple color preselection to efficiently remove low-redshift galaxies and stars \citep{VIPERS_PDR1_2014, VIPERS_PDR2_2018}. This strategy yields high sampling ($\simeq 45$--50\% of all galaxies in its redshift range) over an effective area of about $16.3~\mathrm{deg}^2$ (23.5~deg$^2$ including masked regions), corresponding to a comoving volume of $\sim 4\times 10^{7}\,\mathrm{h}^{-3}\,\mathrm{Mpc}^{3}$ between $z=0.5$ and $z=1.2$ \citep{VIPERS_PDR1_2014, VIPERS_PDR2_2018}.

A primary motivation for VIPERS was to measure galaxy clustering and redshift-space distortions at $z \sim 0.8$ with a sampling density and volume comparable to local surveys, to constrain the growth of structure, test gravity on cosmological scales, and characterize the halo occupation of galaxies \citep{VIPERS_PDR1_2014}. At the same time, the combination of accurate redshifts with extensive multiwavelength photometry has enabled a broad range of galaxy-evolution studies, including measurements of luminosity and stellar-mass functions, environmental trends in galaxy properties, etc \citep{VIPERS_PDR1_2014, VIPERS_PDR2_2018}. Moreover, VIPERS has been extensively used to study passive galaxies, including the development of various sample selection methods and the characterization of properties of massive quiescent galaxies \citep[for e.g.,][]{VIPERS_Fritz_2014_passive, VIPERS_Malavasi_2016_passive, Siudek2016, VIPERS_Garguilo_2017_passive, Siudek_2018_a, Siudek_2018_b}. Therefore, VIPERS provides a well-tested framework and dataset for studying the evolution of massive and passive galaxies at intermediate redshift. Apart from this, VIPERS also targets the general galaxy population at $0.5 < z < 1.2$, rather than targeting specific galaxy samples like DESI \citep{DESI_2016}. So, in contrast to the DESI early data release \citep{DESI_2024}, VIPERS contains more passive galaxies and is a more suitable sample for CC studies. This is the main reason why we chose VIPERS.

In this study, we use the full spectroscopic information from the VIPERS PDR2\footnote{\href{http://www.vipers.inaf.it/}{http://www.vipers.inaf.it/}}, which comprises redshifts, one-dimensional spectra, CFHTLS photometry, and completeness information for 86\,775 galaxies plus 4\,732 additional objects (stars, AGN candidates, and serendipitous sources) \citep{VIPERS_PDR2_2018}. Targets are selected from the CFHTLS-Wide catalogs in W1 and W4 down to an extinction-corrected limit of $i_{\mathrm{mag}} \le 22.5$, with a simple color--color cut in the $(u-g,\,r-i)$ plane to remove galaxies at $z \lesssim 0.5$ and a morphological/SED-based star--galaxy separation that reduces stellar contamination to a few percent \citep{VIPERS_PDR1_2014, VIPERS_PDR2_2018}. Spectroscopic observations were obtained with VIMOS on the ESO VLT, using $1''$-wide slits and the LR-Red grism, which delivers a spectral resolution $R \simeq 220$ over an observed wavelength range of $5500$--$9500~\text{\AA}$ \citep{VIPERS_PDR2_2018}.

\begin{figure}
\includegraphics[clip,width=\columnwidth]{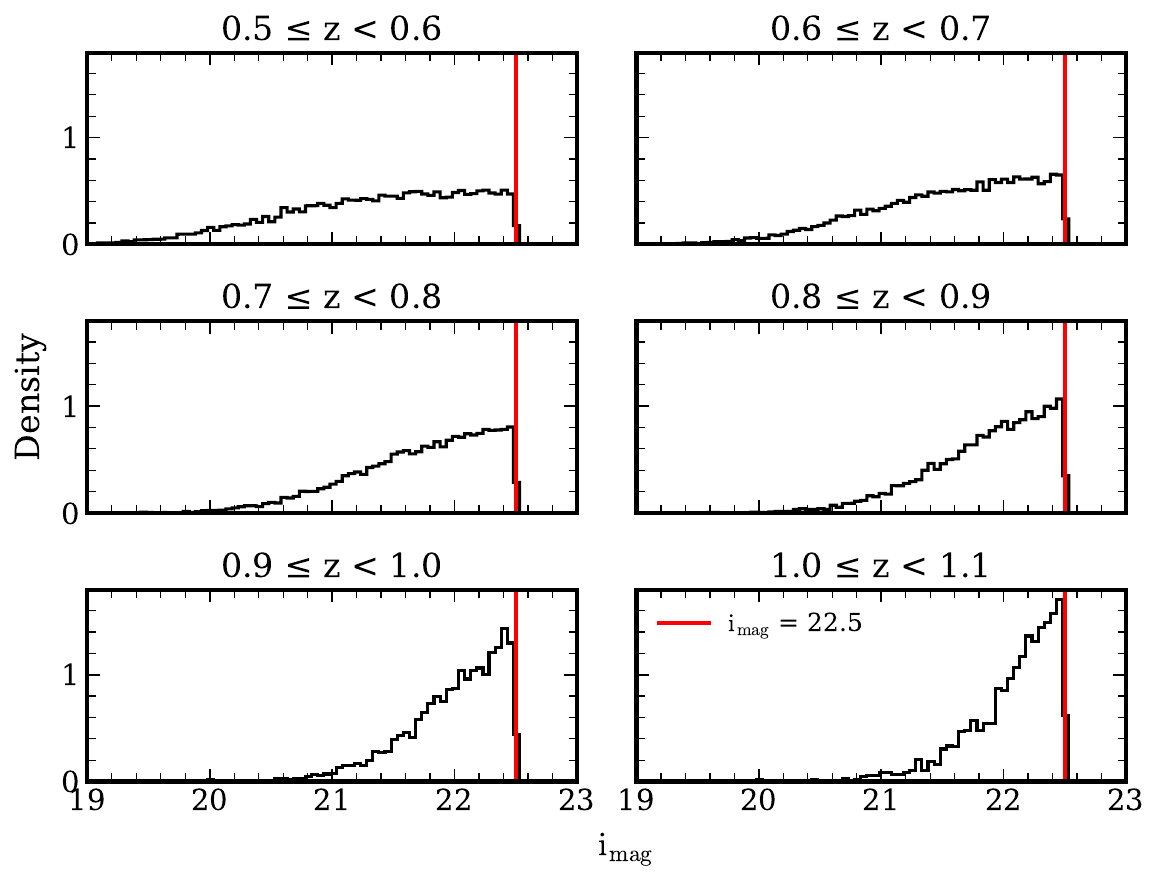} 
\caption{VIPERS $i$-band magnitudes in six different redshift bins of width $dz$ = 0.1. The vertical line represents the magnitude limit of $i_{\rm mag} = 22.5$.}
\label{fig:magcomp}
\end{figure}

For our CC work, using the parent PDR2 sample of 91\,507 objects, we construct a subsample of VIPERS galaxies in the redshift interval $0.5 \le z \le 0.8$ (50\,876 objects), where the survey is highly complete, and the spectral coverage around the 4000~\AA\ break is optimal for cosmic-chronometer applications. We also use the VIPERS spectroscopic redshifts, $z$spec, for our study. In Fig.~\ref{fig:magcomp}, looking at observed $i$ band magnitudes, when we move toward higher redshifts, we reach the magnitude completion limit of VIPERS ($i_{\rm mag} \le 22.5$). This is clearly depicted by the peak density near the vertical line, a marginal increase within $0.5 \le z \le 0.8$, and an abrupt rise thereafter. We remove galaxies at $z > 0.8$ to prevent the magnitude limit from significantly biasing our sample. As redshift increases, an increasing fraction of passive galaxies that would still qualify as CCs may be fainter than the magnitude limit $i_{\rm AB} \leq 22.5$. Therefore, the magnitude cut will remove these galaxies, artificially biasing our sample toward more massive galaxies. Following the downsizing scenario \citep{Neistein_2006_downsizing}, this will result in a bias toward older galaxies, in turn biasing the CCs' differential age measurements. To mitigate such redshift inhomogeneities, some CC studies split their samples into distinct mass or velocity-dispersion bins, in some cases discarding lower-mass or lower-velocity-dispersion bins entirely, to reduce the effect of progenitor bias and downsizing across redshift \citep[e.g.,][]{Moresco_2012b, Borghi_2022a, Tomasetti_2023_spec_cut, Tomasetti_2026, Loubser_2025b}. We do not adopt this strategy here, since our sample size and selected redshift range do not allow a further mass/velocity-dispersion split without significantly reducing the number of galaxies and hence the statistical power.
Within our subsample, we retain only objects with reliable spectroscopic redshifts ($2 \leq z{\rm flg} < 10$; 46\,008 objects), and this selected redshift range already excludes most of the sources classified in the VIPERS catalogs as stars or broad-line AGN. We also have rest-frame VIPERS NUV, r, and Ks magnitudes and stellar masses from SED fitting, as well as D4000 measurements from spectroscopy. The VIPERS galaxies' absolute magnitudes and stellar masses were derived as described in \cite{Moutard_2016b} while fixing $z = z \mathrm{spec} = z_{\rm VIPERS}$ and fitting the VIPERS-Multi Lambda Survey (MLS) photometry \citep{Moutard_2016a}.

The resulting sample is a clean set of galaxies in the desired redshift range and constitutes the parent spectroscopic sample from which we later define our chronometer subsample of passively evolving galaxies (Section~\ref{S:3}). In addition to the spectroscopic redshift measurements, we also use the flux-calibrated VIPERS spectra provided in PDR2 to derive a further refined sample (see Appendix~\ref{Appendix:A}).

\section{Sample selection}
\label{S:3}

In early CC literature \citep[such as][]{Moresco_2012b}{}, a combination of morphological cuts (when available), mass cuts ($\log(M/M_\odot) > 11 \text{ or} > 10.6$ depending on the sample size) and spectroscopic diagnostics was used. They inspected the spectra for emission lines such as H$_\alpha$ and O II (3727\,\AA) to exclude star formation or AGN activity. \cite{Moresco_2013} further analyzed the overlap between different selection methods, including morphological cuts, color-mass cuts, color-color cuts, SED-fitting red galaxies, $\rm sSFR$ cuts, and, finally, their passive ETG sample (a combination of morphological, photometric, and spectroscopic selections). Their final passive ETG sample had only $\sim 30-50\%$ overlap with all other samples, and they also found that when using either color-color or color-mass cuts, $\sim 40\%$ of the sample still had a higher star formation activity (only $\sim 60\%$ overlap with the $\rm sSFR$ cut sample). However, their Fig.~1 (bottom panel) also shows that in higher mass regimes (e.g., $\log(M/M_\odot) \geq 11$) the overlap with the sSFR cut sample is considerably higher, meaning that a high mass cut increases the fraction of quiescent galaxies significantly. 

Until now, studies have used high-quality spectroscopic samples and focused on carefully validating their passive galaxy sample by visual inspection of the spectra for emission lines and also by excluding galaxies that have an equivalent width (EW) of O II and O III lines more than 5\,\AA\, \citep[e.g.,][]{Borghi_2022a, Loubser_2025b}{}. Past literature has also proposed and tested the reliability of using the ratio of minimum fluxes in Ca II H and Ca II K (or the ratio of two pseudo Lick indices as proposed by \cite{Borghi_2022a}) as a diagnostic of passivity, and it was found that Ca II H/K $\lesssim$ 1.1 characterizes passive galaxies \citep{Moresco_2018, Borghi_2022a}. Moreover, the D4000 index itself is a very good indicator of passivity and has been used to refine passive galaxy samples \citep[e.g., see][]{Figueira_2024}. However, in the D4000-based CC framework, we use the D4000 index as a proxy for age estimation, and including it in the sample selection would introduce intrinsic bias to our sample. As a standard procedure, recent CC studies now select only passive galaxies using a combination of photometric (color-color, stellar mass cuts, or velocity dispersion bins) and spectroscopic diagnostics (EWs of O II, O III, and Ca II H/K). However, since we are preparing our framework for a photometry-only study, we explore whether we can obtain reliable $H(z)$ measurements solely using photometric sample selection.
For the sake of completeness, we also explore the effect of a more conventional spectroscopic sample selection in Appendix~\ref{Appendix:A}.

\subsection{Color cuts}
\label{SS:3.1}

For color-color diagrams, we use the $(NUV-r)$ and $(r-K)$ colors, which were proposed by \cite{Arnouts_2013}, based on the color-color diagrams of \cite{Williams_2009}. Here, we directly use a modified version of the color cuts for the VIPERS sample (PDR1) proposed by \cite{Moutard_2016b}. Following the low-density green valley of the NUVrK diagram, they define the color cut as
\begin{equation}\label{eq:Moutard_cc1}
    [(NUV-r) > B_2] \cap [(NUV-r) > A\,(r-K) + B_1]\,,
\end{equation}
where $A$ is the slope, which was found to be $\sim 2.25$ for all redshift bins, and $B_1$ and $B_2$ are redshift-dependent parameters that evolve as a function of the lookback time ($t_{\rm L}$) \citep{Moutard_2016b}. They first trace the green valley boundaries and make a color cut such that it runs right through it.

The dataset used in our work is a subsample of the VIPERS dataset used by \cite{Moutard_2016b}, so we expect that a stringent version of their color cuts also works for us. Our goal is to shift the color cuts upward to extract a sufficiently pure passive sample regardless of the sample completeness. So, we keep the slope the same ($A = 2.25$) and tweak the other two parameters, $B_1$ and $B_2$. Using Fig.~7 from \cite{Moutard_2016b}, we select the upper envelope of the green valley to define $B_1$: $B_1(t_{\rm L}) = -0.029\, t_{\rm L} + c_0$, where $c_0$ is the intercept of the upper green valley boundary and is $\sim 3$. We empirically set $B_2(t_{\rm L} = 5\, \mathrm{Gyr}) = (NUV-r_{\rm h0}) - 0.029\, t_{\rm L} = 4.5 - 0.145 = 4.355$, where $(NUV-r_{\rm h0})$ is the initial horizontal intercept. This choice stringently removes trails of star-forming and transitioning galaxies, although we might have lost a small number of passive galaxies. Our final color-cut can be written as
\begin{equation}\label{eq:Pradhan_cc}
\begin{aligned}
    [(NUV-r) > (NUV-r_{\rm h0}) - 0.029\, t_{\rm L}] \\
    \,\cap\, [(NUV-r) > 2.25\,(r-K) - 0.029\, t_{\rm L} + c_0]\,,
\end{aligned}
\end{equation}
with $(NUV-r_{\rm h0}) = 4.5$ and $c_0 = 3.0$.

\begin{figure} 
    \centering
    \begin{subfigure}{\columnwidth}
        \includegraphics[width=\columnwidth]{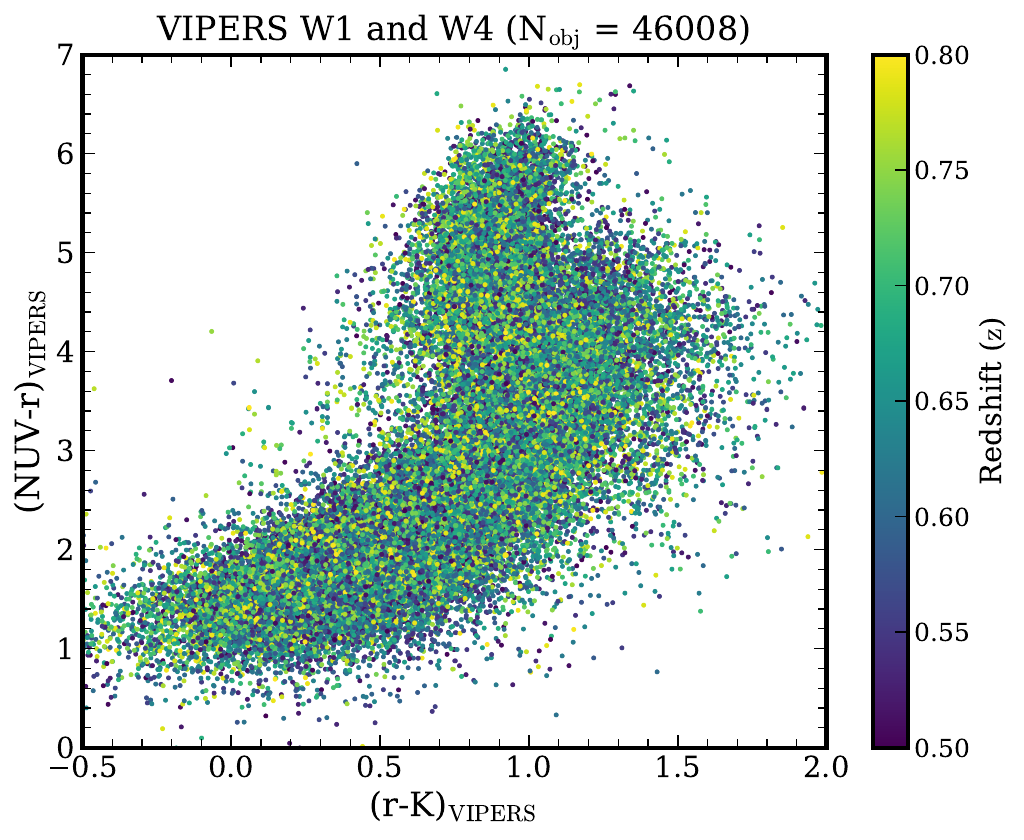}
    \end{subfigure}
    \begin{subfigure}{\columnwidth}
        \includegraphics[width=\columnwidth]{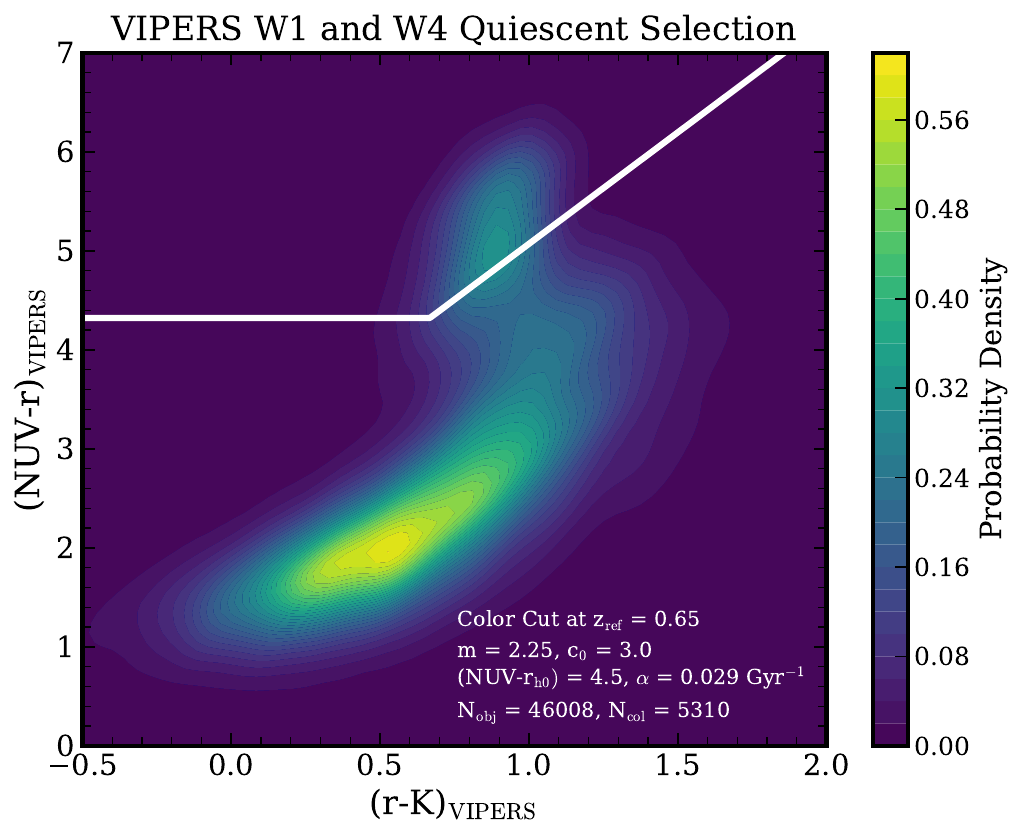}
    \end{subfigure}
    \caption{VIPERS NUVrK color-color diagram for W1 and W4 fields combined colored by redshift (top panel). KDE plot of NUVrK diagram with a modified version of the color cut prescribed in \protect\cite{Moutard_2016b} (bottom panel).}
    \label{fig:colorcut}
\end{figure}

This color cut is shown in the bottom panel of Fig.~\ref{fig:colorcut}, overlaid on a kernel density plot. Each level, from darker to lighter colors, depicts the increasing probability density of galaxies, and we can clearly see how it separates the red and blue clouds. The color cut passes through the red cloud of passive galaxies, removing the galaxies in the green valley. Note that the color cut line only represents the cut at a reference redshift of $z_{\rm ref} = 0.65$, but in the final color cut sample, Eq.~\eqref{eq:Pradhan_cc} has been used to classify individual galaxies at their respective redshifts. 
Using this color cut, we selected $N_{\rm col} = 5\,310$ ($\sim 11.54\%$) galaxies out of $N_{\rm obj} = 46\,008$ objects in total. 

We note that the redshift dependence of this color cut reflects the global aging of the galaxy population \citep{Moutard_2016b}, derived independently of any CC observable. This redshift-dependent cut is specifically designed to keep selecting the same type of galaxy (quiescent, by a consistent definition) at every epoch, which is what avoids a redshift-dependent selection bias. Additionally, in Section~\ref{SS:3.2} we combine this color cut with a fixed stellar-mass threshold that does not evolve with redshift, decoupling our final selection from the D4000-based age trend measured in Section~\ref{S:5}.

\subsection{Mass cuts}\label{SS:3.2}
Color cuts alone are not enough, and as we know from the earlier discussions, our color-selected sample still has impurities. Without spectroscopic selection, we cannot get a 100\% pure sample, but doing a stringent mass cut would improve the overlap of our sample with a pure spectroscopically selected sample (see Fig.~\ref{fig:P_passive_hist_mcut}). However, the main reason behind using a mass cut for CC studies is to prevent bias from the mass downsizing scenario: studies of stellar populations show that massive ETGs form their stars earlier and on shorter timescales, while less massive ones continue to form stars for longer periods \citep{Neistein_2006_downsizing, Gallazzi_2006_downsizing}. Therefore, even in a passive sample, if there is a mix of low and high masses, the galaxies with extended SFHs and younger components would bias the ages to lower values since the galaxies in a particular redshift bin are not coeval \citep{Moresco_2012b, Loubser_2025b}. Furthermore, massive and passive galaxies also have solar or mildly super-solar metallicities with very small intrinsic scatter \citep[see][]{Gallazzi_2006_downsizing, Moresco_2012b, Loubser_2025b}, which means that selecting a more massive sample narrows the age--metallicity parameter space probed by D4000$_n$ (see Fig.~\ref{fig:d4000_age}), thereby reducing residual uncertainties from D4000-age-metallicity degeneracy.

Earlier CC studies, such as \cite{Moresco_2012b}, use mass cuts of $\log(M/M_\odot) \geq 10.6 \text{ or } 11$ and \cite{Moresco_2016} uses $\log(M/M_\odot) \geq 10.75$, which was also followed by \cite{Loubser_2025b}. Moreover, \cite{Veale_2017} notes that galaxies above $\log(M/M_\odot) = 10.5$ are generally slow rotators and passive \citep{Loubser_2025b}.

\begin{figure} 
    \centering
    \begin{subfigure}{\columnwidth}
        \includegraphics[width=\columnwidth]
        {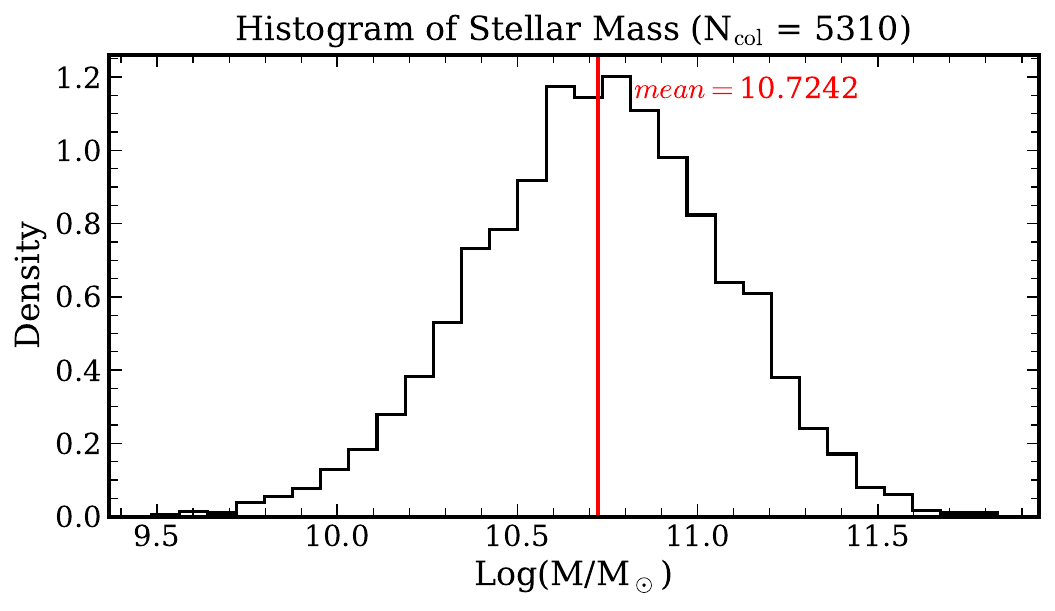}
    \end{subfigure}
    \begin{subfigure}{\columnwidth}
        \hspace{0.018\columnwidth}
        \includegraphics[width=\columnwidth]
        {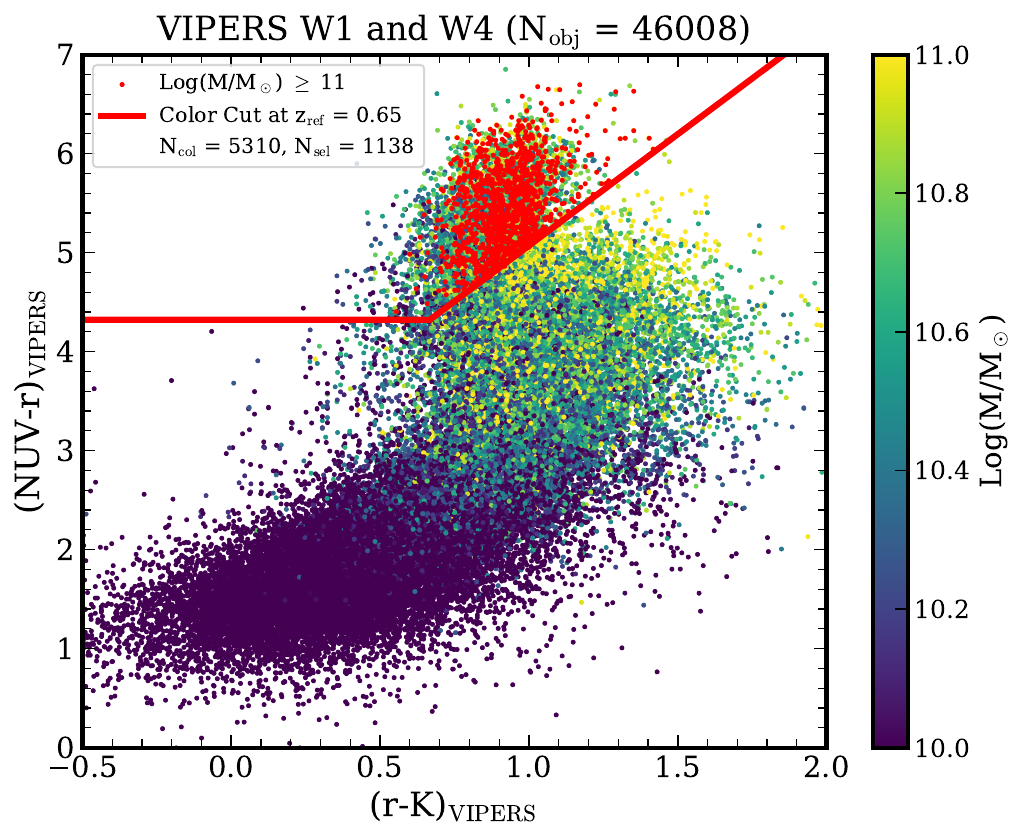}
    \end{subfigure}
    \caption{VIPERS mass histogram of the color-cut sample (top panel). NUVrK scatter plot colored by mass, with the color cut shown in red and the intersection of color and Log(M/M$_\odot$) $\geq$ 11 shown as red dots (bottom panel).}
    \label{fig:masscut}
\end{figure}

To motivate our mass threshold, we examine the mass histogram of the color cut sample in Fig.~\ref{fig:masscut} top panel. We find the mean mass to be $\log(M/M_{\odot}) \simeq 10.72$, and there is a significant population of galaxies above it. In the bottom panel (NUVrK scatter plot colored by mass), we also see that the most massive galaxies occupy redder regions of the diagram, which provides a secondary check for passivity.

Since we do not use spectroscopic indicators (such as EW of emission lines) for further selection, unlike the previously mentioned studies, we do a more stringent mass cut of $\log(M/M_\odot) \geq 11$. The galaxies selected this way are shown as red dots in the bottom panel of Fig.~\ref{fig:masscut}. Finally, we have a color- and mass-selected sample of $N_{\rm sel} = 1138$ galaxies, which is only $\sim 2.47\%$ of the original sample.

\subsection{Final sample for cosmic chronometers}
\label{SS:3.3}

\begin{figure}
\includegraphics[clip,width=\columnwidth]{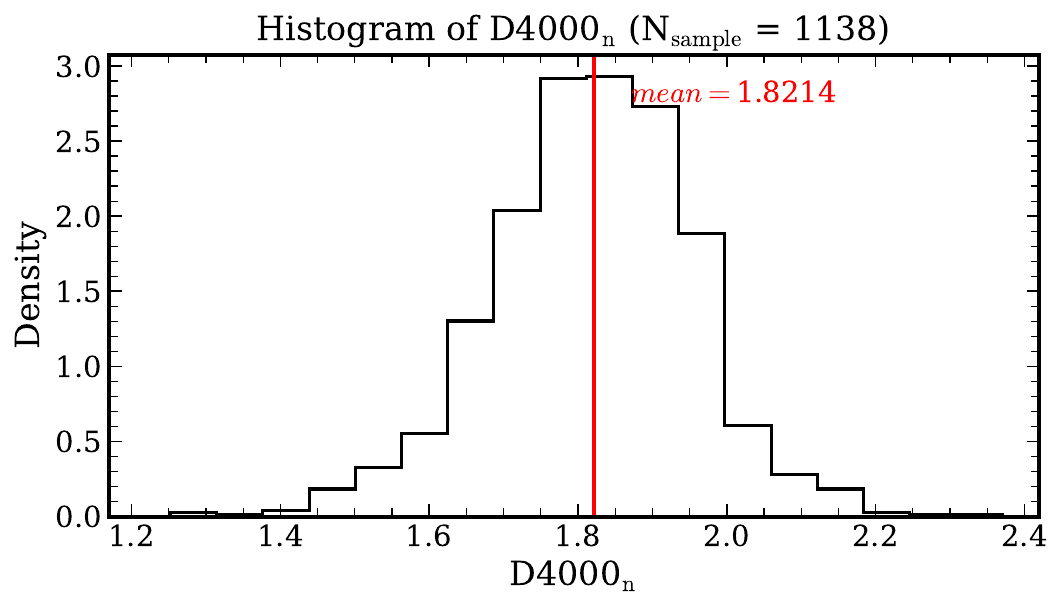} 
\caption{D4000$_n$ histogram of the color- and mass-selected passive sample. The vertical line denotes the mean of the distribution.}
\label{fig:d4000_hist}
\end{figure}

We now have our final sample of 1138 galaxies for use as chronometers. These galaxies also have reliable $\mathrm{D}4000_n$ measurements with signal-to-noise ratio (S/N) $> 3$ in the range of 1.4 to 2.2 (alongside some outliers beyond this range), as depicted in Fig.~\ref{fig:d4000_hist}. The mean $\mathrm{D}4000_n$ value of $\sim 1.82$ further testifies that most of the galaxies in our sample are passive. Having lower $\mathrm{D}4000_n$ is also not uncommon in existing CC studies \citep[see e.g.,][]{Moresco_2012b, Loubser_2025b} which generally have a Gaussian distribution of passive galaxies with most galaxies in the central $\mathrm{D}4000_n$ range of 1.6 to 2. In that regard, our sample shows a similar trend. Our sample is not the largest compared to works such as \cite{Moresco_2012b} and \cite{Loubser_2025b}, but it is modest compared to many others \citep[e.g.,][]{Borghi_2022a, Borghi_2022b}. 

\subsection{Spectroscopic sample selection}
As a control test (different from our fiducial photometry-based sample selection), we additionally perform a conventional spectroscopic sample selection on our color- and mass-selected spectro-photometric sample\footnote{Note that $z$spec has been used for SED fitting VIPERS data, and since our colors and masses are derived from this, our final sample is not purely photometric.}, in order to quantify how much a standard spectroscopic passivity cut (as used in prior CC literature) would change our sample and resulting H(z) constraints relative to the purely photometric selection adopted in Section~\ref{SS:3.1}--\ref{SS:3.3}.
For this, we use two standard indicators, viz., the EW of the O II emission line \citep[$< 5\, \text{\AA}$;][]{Moresco_2012b, Loubser_2025b} and the Ca II H/K ratio ($> 1.1$), both measured following the prescription in \cite{Borghi_2022a} using their \texttt{PyLick} code. Using a Bayesian framework to assign a probability of being passive to each galaxy, we create a pure subsample of galaxies with a probability $\geq 0.9$ of being passive. Subsequently, we test using the hyper-pure sample to determine whether low-to-moderate-resolution spectroscopy in VIPERS can be reliably used to measure $H(z)$ compared to measurements from photometric sample selection. We find that the improvement in the sample mean D4000 is marginal, and the $H(z)$ constraints are not significantly altered. Please see Appendix~\ref{Appendix:A} for more details.

\section{Stellar population models}
\label{S:4}

Now that our sample is ready, we need to infer the age differentials in our sample using D4000-Age relations from SPS models. Earlier CC studies \citep[such as][]{Moresco_2011, Moresco_2012b} used these relations from SPS models to estimate the slope ($\rm A(Z, SPS)$ parameter), which was then plugged directly into the CC equation. 
They used an exponentially declining star formation rate, $\mathrm{SFR}(t) \propto t/\tau^2\, exp(-t/\tau)$, where $\tau$ is the star formation timescale. \cite{Moresco_2011} found from the analysis of an SDSS ETG sample that $\tau$ has a median value of less than 0.2~$\mathrm{Gyr}$ for all mass subsamples and for the majority of the SEDs, $\tau \le 0.3$ was required, which is also consistent with other ETG studies \citep{Moresco_2012b}. 

Furthermore, as mentioned in Section~\ref{S:1}, \cite{Moresco_2020} studied the effects of different SPS models, their parameters (such as $\tau$), IMFs, and stellar libraries. They quantified these effects in the form of a covariance matrix, and the corresponding systematic errors can be added directly to the age and H(z) measurements in future ETG and CC studies. One recent study that successfully used this kind of covariance matrix to propagate errors is \cite{Loubser_2025b}. Note that in all these methods, the systematics are condensed into $\rm A(Z, SPS)$. In contrast, we directly derive stellar ages from the SPS model grids corresponding to our D4000 measurements.

To create our SPS grids, we use \cite{Siudek2016} and \cite{Moresco_2012b} as references, both of which used these to study ETGs, and the former used it for VIPERS. We use the \texttt{PyGALAXEV}\footnote{\href{https://github.com/astrosonnen/pygalaxev}{https://github.com/astrosonnen/pygalaxev}} code, which is a Python wrapper of the \texttt{GALAXEV}\footnote{\href{https://www.bruzual.org/bc03/}{https://www.bruzual.org/bc03/}} stellar population synthesis code developed by \cite{BC03} and is used to predict broad-band magnitudes on a grid of SPS parameters. This code uses the \texttt{BC03} population synthesis models \citep{BC03} to generate synthetic spectra, using \cite{Bertelli_1994_Padova_Met_Tracks} stellar evolutionary tracks and covering a wide range of metallicities. We also use the high-resolution STELIB spectral library with an exponentially decaying SFH having $\tau = 0.1~\mathrm{Gyr}$ and assuming there is little to no dust extinction in ETGs \citep[see][]{Siudek2016}. With these parameters set, we obtain synthetic spectra for ages ranging from 1 to 14~$\mathrm{Gyr}$, with a grid size of 560 points.

Our goal here is limited only to deriving D4000--age--metallicity relationships, and for this, we customize the code to calculate D4000 values from the models' spectra as
\begin{align}\label{eq:D4000}
     \mathrm{D}4000 = \frac{\dfrac{1}{\Delta \lambda_{\rm red}}\displaystyle\int_{\lambda_{\rm red}}^{\lambda_{\rm red} + \Delta \lambda_{\rm red}} F_\lambda \, d\lambda}{\dfrac{1}{\Delta \lambda_{\rm blue}}\displaystyle\int_{\lambda_{\rm blue}}^{\lambda_{\rm blue} + \Delta \lambda_{\rm blue}} F_\lambda \, d\lambda}\,,
\end{align}
where $F$ is the flux, $\lambda_{\rm red}$ and $\lambda_{\rm blue}$ are the red and blue continuum limits, and the corresponding $\Delta \lambda_{\rm red}$ and $\Delta \lambda_{\rm blue}$ are the widths of their continuum windows. The red and blue continuum ranges are presented in Table~\ref{tab:d4000_bands} for the wide (D4000$_w$) and narrow (D4000$_n$) definitions of the D4000 index. However, we only use the D4000$_n$ values following existing CC literature, since: (i) it is less sensitive to dust reddening \citep{Balogh_1999}; (ii) it is the standard for CC studies due to its higher age sensitivity \citep{Moresco_2011, Moresco_2012a, Moresco_2012b, Loubser_2025b}; and (iii) it needs a narrower wavelength range to be fully sampled, especially advantageous at $z \gtrsim 0.5$ where S/N degrades.

\begin{table}
    \centering
    \caption{Wavelength ranges for the two definitions of the D4000 index.}
    \label{tab:d4000_bands}
    \begin{tabular}{lcc}
        \hline \hline
        \textbf{Index} & \textbf{Blue band (\AA)} & \textbf{Red band (\AA)} \\
        \hline
        $\mathrm{D}4000_w$ \citep{Bruzual_1983}   & 3750--3950 & 4050--4250 \\
        $\mathrm{D}4000_n$ \citep{Balogh_1999} & 3850--3950 & 4000--4100 \\
        \hline
    \end{tabular}
\end{table}

Finally, we obtain $\mathrm{D}4000_n$--stellar age relationships based on the above-mentioned measurement criteria, from \texttt{BC03} models with metallicities, $\log(Z/Z_\odot) = 0.40$, $\log(Z/Z_\odot) = 0$, $\log(Z/Z_\odot) = -0.40$, and $\log(Z/Z_\odot) = -0.70$. The model grids can be written as $\mathrm{D}_{\mathrm{model}}(t,Z)$, where $\mathrm{D}$ represents $\mathrm{D}4000_n$ value, and $t$ and $Z$ represent the age and metallicity, respectively. These are depicted in Fig.~\ref{fig:d4000_age}, where we clearly notice how higher D4000$_n$ values correspond to higher ages, and for a given age, higher metallicities correspond to higher D4000 values. This figure can be compared with Fig.~4 of \cite{Moresco_2012b} and Fig.~9 of \cite{Siudek2016} to verify how similar they are. We use these grids to measure ages for the VIPERS galaxies in the next section. Please also note that, unlike some other studies, we do not examine the effects of different SFH parameters (e.g., different $\tau$ values), SPS models, or stellar libraries. These effects are extensively studied in the CC literature, and we use their results to derive the systematic errors arising from these effects \citep[e.g.,][]{Moresco_2020}.

\begin{figure}
\includegraphics[clip,width=\columnwidth]{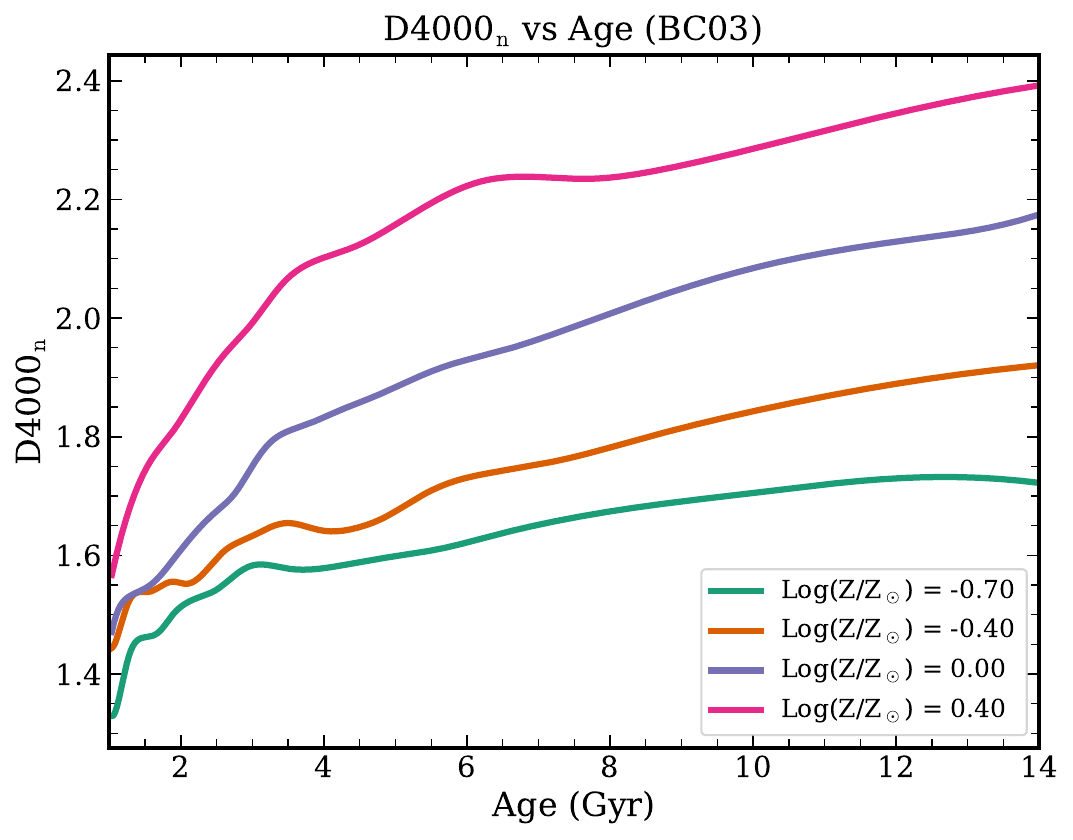} 
\caption{D4000$_n$ vs. age from \texttt{BC03} stellar population synthesis models for different metallicty bins. This plot represents the SPS grid used for age calculations.}
\label{fig:spsmodel}
\end{figure}

\section{Age estimation}
\label{S:5}

With the passive galaxy sample ready, containing D4000$_n$ and redshift data, and the SPS model grids from \texttt{BC03} containing D4000$_n$--age--metallicity information, we are ready to estimate the ages of the CC galaxy sample. The resulting D4000$_n$ distribution is shown in Fig.~\ref{fig:d4000_age} (top panel), where darker-colored boxes represent a higher density of galaxies and vice versa. The decreasing trend of D4000$_n$ with redshift is clear, and most of the galaxies are concentrated in $1.6 < \mathrm{D}4000_n < 2$. Furthermore, these trends are consistent with the standard CC literature \citep[see e.g.,][]{Moresco_2012b, Loubser_2025b}.

Before deriving the ages, we first bin the D4000$_n$ measurements in redshift and consider only the median D4000$_n$ measurements. For each bin, we define the error of the median as the median absolute deviation (MAD) divided by the square root of the number of objects in the bin ($\sigma_{\rm D} = \text{MAD}/\sqrt{n}$). Standard CC literature following the D4000 method of \cite{Moresco_2011} also follows the same method of binning and taking the median of the bin with the MAD \citep[e.g.,][]{Moresco_2012b, Loubser_2025b}. They did this because the median is much less sensitive to outliers (such as residual trends with mass, color, or subtle contamination by residual young components) than the mean, so it better represents the ``typical'' passive ETGs in that bin. The distribution of D4000 values within a bin is often skewed or has extended tails (e.g., because of residual star formation or metallicity spread), so using the median and the MAD gives an uncertainty estimate that is robust to non-Gaussian scatter, rather than relying on the standard deviation, which is very sensitive to tails. We do not want the error bars to be dominated by a small number of problematic galaxies and therefore we follow the same method. However, the next question that naturally follows is: how to bin the D4000$_n$ distribution in redshift such that most of the bins are consistent with their neighboring bins, further resulting in a consistent decreasing trend with redshift as expected?

\subsection{Binning strategies}
\label{SS:5.1}

\begin{figure}
    \centering
    \begin{subfigure}{\columnwidth}
        \includegraphics[width=\columnwidth]{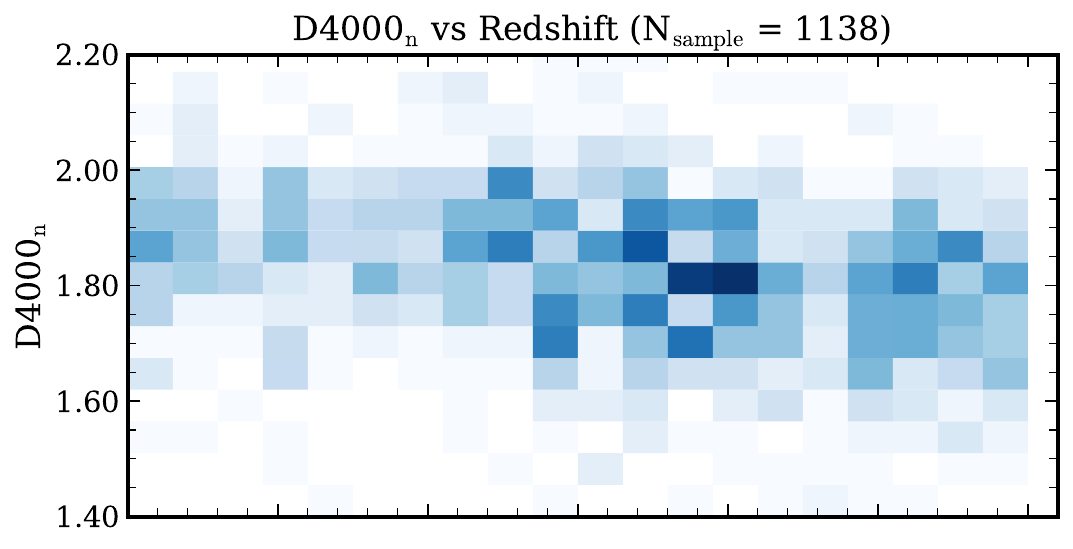}
    \end{subfigure}
    \begin{subfigure}{\columnwidth}
        \includegraphics[width=\columnwidth]{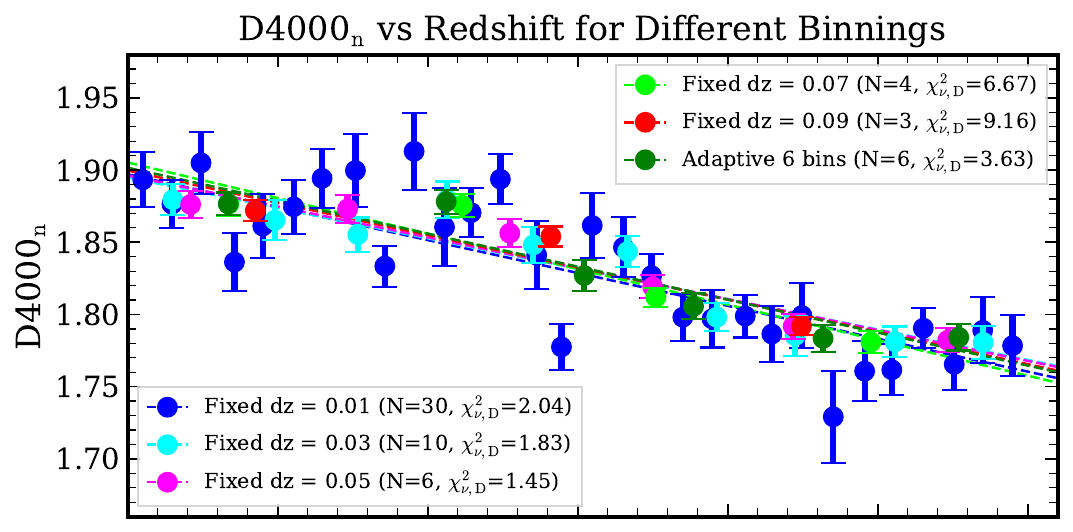}
    \end{subfigure}
    \begin{subfigure}{\columnwidth}
        \includegraphics[width=\columnwidth]{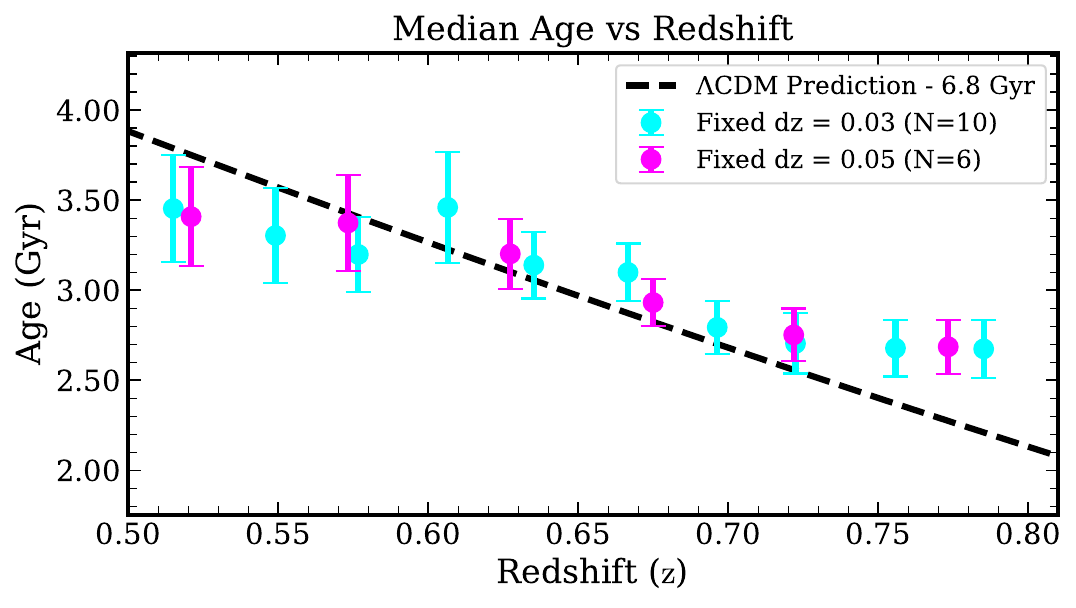}
    \end{subfigure}
    \caption{Plot of D4000$_n$ distribution (top panel), D4000$_n$ vs. redshift for different binning strategies (middle panel), age vs. metallicity fitted with $\Lambda$CDM prediction for the selected binnings (bottom panel; median ages and $\sigma_{68}$).}
    \label{fig:d4000_age}
\end{figure}

To answer the aforementioned question, we test six different binning strategies: (i) five redshift binnings with bin widths, $dz$ = 0.01, 0.03, 0.05, 0.07, and 0.09 (resulting in N= 30, 10... bins respectively); and (ii) one adaptive redshift binning with a total of six bins containing almost the same number of objects ($\sim 191$/bin). Please note that for fixed-width redshift binning with bin widths, $\textrm{d}z > 0.5$, some data points toward the end of the redshift range get clipped off since we do not have enough redshift range for the larger $dz$ to cover. For example, when we set $dz$ = 0.07, we get four bins up to 0.78, and after that, the boundary at $ z$ = 0.8 is only $dz$ = 0.2 away. We make a scatter plot with all six binning strategies and fit each with a straight line, as seen in Fig.~\ref{fig:d4000_age} (middle panel). We can clearly see significant scatter across most binning choices, and some points appear to be outliers. However, it is difficult to visually confirm which points are outliers and which binning strategies have excessive scatter. 

To resolve the outlier issue, we performed 3-sigma clipping relative to the fit line and found that all points are within 3-sigma. Furthermore, to quantify the scatter \citep{Moresco_2012b, Loubser_2025b}, we measure the reduced $\chi^2$ ($\chi_\nu^2$, which is simply the $\chi^2$ per degree of freedom (d.o.f)), with the fitting line as the reference. $\chi_\nu^2$ values $\sim 1$ imply that the binned D4000 values are very consistent with a linear trend, and thus no D4000 values are significantly offset from adjacent bins. These values are mentioned in the legends of Fig.~\ref{fig:d4000_age} (middle panel), for each of the binning strategies as $\chi^2_{\nu,\mathrm{D}}$\footnote{Here, the index D means D4000 and has been used to attribute the corresponding scatter measurements to the D4000 distribution while also differentiating it from the other $\chi_\nu^2$ values we present later.}. For further analysis, we select only two of the best binning strategies with the least scatter: $dz$ = 0.03 and 0.05, with $\chi_{\nu,\mathrm{D}}^2$ = 1.83 and 1.45, respectively. Note that for these two binning strategies, all points were found to lie within 3$\sigma$; consequently, no points were removed and the full binned dataset was retained for subsequent analysis.

\subsection{Bayesian framework}
\label{SS:5.2}

In this work, we do not propagate median D4000 errors and metallicity uncertainties independently. Instead, we use Bayesian inference to account for both simultaneously, sampling a 2D probability space. The goal of this more rigorous statistical treatment is to apply it to photometry-only CCs in upcoming work, where both observational errors and metallicity uncertainties may be larger. The key idea here is to treat the observed median $\mathrm{D}4000_n$ index in each redshift bin, $\mathrm{D}$, as a noisy realisation of a forward SPS model $\mathrm{D}_{\rm model}(t, Z)$ with the model parameters $t$ and $Z$ representing the age and metallicity of the passive galaxy population in that bin respectively. Then, using Bayes' theorem, we get a probability distribution $P(t, Z\mid \mathrm{D})\propto P(\mathrm{D}\mid t, Z)P(t)P(Z)$, which can be marginalized over metallicity to obtain the age posterior $P(t\mid \mathrm{D})$.

The detailed construction of the likelihood and the numerical implementation of the marginalization procedure for age estimation on a $(t, Z)$ grid are described in Appendix~\ref{Appendix:D_to_age}. In this section, we only outline the conceptual structure and the role of the priors. The likelihood $P(\mathrm{D}\mid t, Z)$ encodes the mapping from SPS grids to observed $\mathrm{D}4000_n$ along with the observational errors on $\mathrm{D}$ folded in. The age prior $P(t)$ is deliberately broad and flat over a physically plausible age range, making sure it does not impose any cosmology-dependent age-redshift relation. In contrast, the metallicity prior $P(Z)$ is informed by independent measurements of the stellar metallicities of the passive galaxies from the literature, and therefore plays an important role in breaking the age-metallicity degeneracy. After marginalization over $Z$, the resulting $P(t\mid \mathrm{D})$ directly reflects the information about stellar age carried by the observed D4000$_n$ under the adopted SPS and metallicity assumptions.

Figure~\ref{fig:prior_posterior} illustrates two key ingredients of our CC framework. The top panel shows the adopted metallicity prior and compares it to discrete metallicity values from previous CC analyses. The bottom panel shows a representative age posterior for one of our binning schemes ($dz$ = 0.05), highlighting both the non-Gaussian shape of the posterior and the way the data and prior combine to constrain the age.

\subsubsection{Metallicity prior}
\label{SSS:5.2.1}

The metallicity prior $P(Z)$ is a key ingredient of our Bayesian marginalization, as D4000$_n$ is sensitive to both age and metallicity. For massive, quiescent galaxies, several studies have found that stellar metallicities are approximately solar or mildly super-solar, with no significant evolution over the redshift range relevant for current CC applications \citep[e.g.,][]{Gallazzi_2014, Onodera_2015, Citro_2016, Estrada-Carpenter_2019, Borghi_2022a, Jiao_2023}. However, this is not the general consensus, and studies such as \cite{Beverage_2021} report a non-negligible $\sim 0.1$--$0.2$ dex evolution for $z \geq 0.7$. To explicitly test how metallicity evolving with redshift might affect our $H(z)$ constraints, we repeated our analysis using a set of redshift-dependent metallicity priors in Appendix~\ref{Appendix:C.1}. Regardless, as a standard practice, CC studies assume a fixed non-evolving metallicity with redshift. Some early CC works, such as \cite{Moresco_2012b}, assume a constant metallicity of $Z/Z_\odot = 1.1 \pm 0.1$ for their $z > 0.3$ sample. Recent work by \cite{Loubser_2025b} sets a fixed mean metallicity equal to solar metallicity and makes a reasonable assumption of 10\% error on this. Some of the metallicity values from the literature mentioned are shown in Fig.~\ref{fig:prior_posterior} (top panel), with the mean shown as a solid line and the error boundaries as dashed lines. Please note that the errors depicted in the figure are root-mean-square (rms) errors and not Gaussian errors. The \cite{Borghi_2022a} prior is broader, as their study focuses on individual galaxies and measures object-to-object dispersion. We adopt a Gaussian prior on the bin-averaged stellar metallicity centered on the mean value measured by \cite{Citro_2016} for massive and passive Sloan Digital Sky Survey (SDSS) galaxies, $\mu_{\rm Z} = 0.027$ and their measurement uncertainty of $\sigma_{\rm Z} = 0.002$ as the dispersion (after converting to dex).
This value is the MS11-based estimate reported by \cite{Citro_2016}; their BC03-based estimate is $Z=0.029\pm0.0015$, differing by only $\Delta Z\sim0.002$. Although our own D4000-age-metallicity grids are constructed using BC03 models (Section~\ref{S:4}), this small discrepancy is well within the SPS-model systematic uncertainty already included in our error budget (Section~\ref{SS:6.1}). Moreover, the difference is an order of magnitude smaller than the metallicity perturbations tested explicitly in Appendix~\ref{Appendix:C}. Within their own analysis, \cite{Citro_2016} compare this value to independent literature measurements for massive, passive galaxies, finding close agreement with \cite{Gallazzi_2005} and \cite{Thomas_2010_Passive} (differences of only $\Delta Z\sim-0.0032$ and $0.0036$, respectively) but a larger offset relative to \cite{McDermid_2015} ($\Delta Z\sim-0.01$), whose values are closer to solar. They note that absolute metallicity values depend on the method used (e.g., full-spectrum fitting versus Lick-index-based estimates), even where studies agree qualitatively on supersolar metallicities ($Z>0.025$).

Comparing our adopted prior against the literature values shown in Fig.~\ref{fig:prior_posterior} (top panel), our prior does not overlap at the $1\sigma$ level with either the solar-metallicity assumption or with the value from \cite{Moresco_2012b} ($Z/Z_\odot=1.1\pm0.1$), but is fully consistent with the broader, dispersion-based prior of Borghi et al. (2022b). This reflects mild ($\lesssim2\sigma$) tension with some earlier solar-metallicity assumptions rather than uniform quantitative agreement across all cited references. However, we note that unlike \cite{Moresco_2012b} and \cite{Loubser_2025b}, who assume a fixed metallicity value, our prior is anchored to a direct observational measurement from $\gtrsim24{,}000$ massive, passive SDSS galaxies \citep{Citro_2016}.

This choice is motivated by two considerations. First, it incorporates
external empirical knowledge about the metallicity of massive ETGs, which is otherwise only weakly constrained by D4000$_n$ alone.
Second, it preserves the cosmology independence of the CC method, since
$P(Z)$ encodes purely astrophysical information and does not depend on any assumed $(\Omega_{m},\Omega_\Lambda,\omega(z))$ or on a model-dependent age-redshift relation. To assess the robustness of our results to this assumption, we repeated the analysis with broadened and shifted metallicity priors (e.g.,\ doubling $\sigma_Z$ or shifting $\mu_Z$ by $\pm 0.05$\, dex). The resulting variations in $H(z)$ remain well within our quoted systematic error budget, indicating that, for our data quality, the inference is predominantly likelihood-dominated rather than prior-driven (see Appendix~\ref{Appendix:C} for more details).

\subsubsection{Age posteriors}
\label{SSS:Age_posteriors}

\begin{figure}
    \centering
    \begin{subfigure}{\columnwidth}
        \includegraphics[width=\columnwidth]{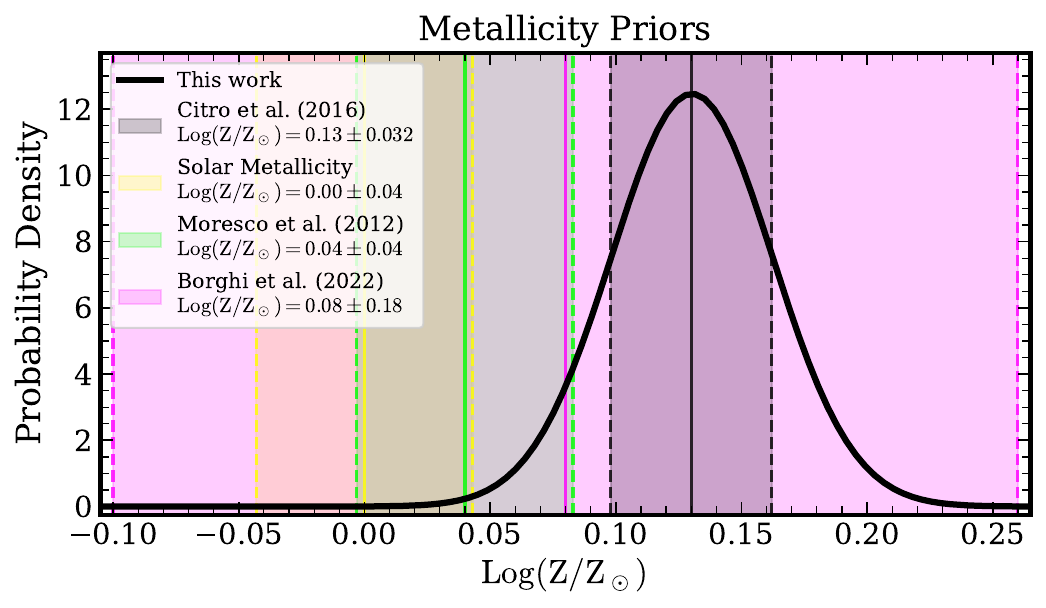}
    \end{subfigure}
    \begin{subfigure}{\columnwidth}
        \includegraphics[width=\columnwidth]{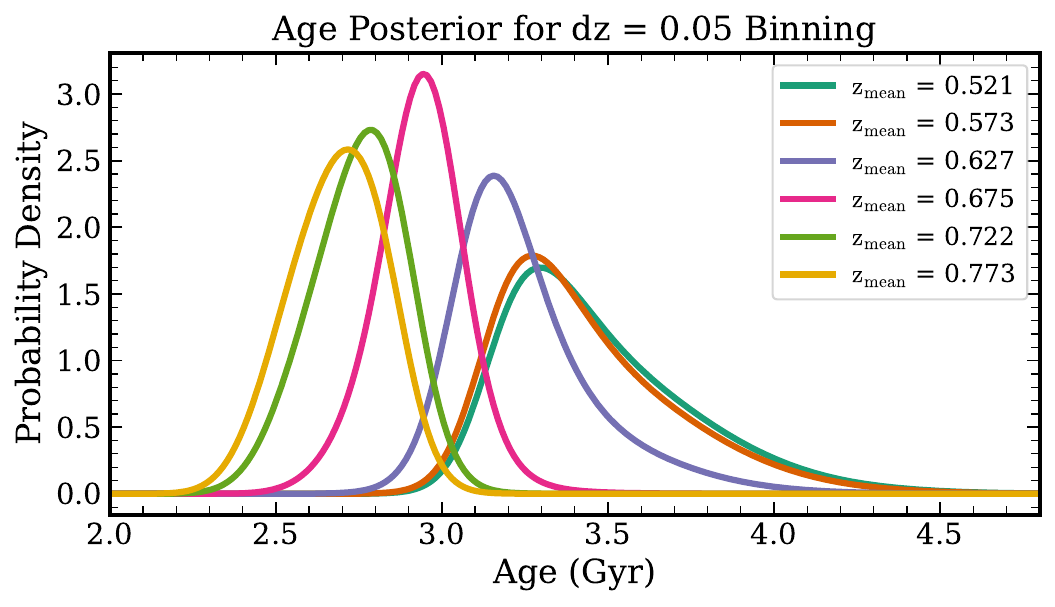}
    \end{subfigure}
    \caption{Plot of metallicity prior used in our study (top panel) compared to solar metallicity and priors (mean and width) available from literature. Age posterior distribution for one of the binning cases ($dz$ = 0.05) (bottom panel).} 
    \label{fig:prior_posterior}
\end{figure}

Given the likelihood and priors, we now evaluate the joint posterior $P(t, Z \mid \mathrm{D})$ on a regular $(t, Z)$ grid and marginalize over $Z$ to obtain the age posterior $P(t \mid \mathrm{D})$ for each redshift bin. The bottom panel of Fig.~\ref{fig:prior_posterior} shows the age posteriors for the $dz$ = 0.05 binning scheme. The posteriors are clearly non-Gaussian, with mild asymmetry around the median. This illustrates the limitations of purely independent error propagation from D4000$_n$ to age and motivates our use of the Bayesian framework to propagate these realistic uncertainties into the final $H(z)$ measurements. We characterise each posterior by its median age $t_{50}$ and the 16$^{\rm th}$ and 84$^{\rm th}$ percentiles $(t_{16},t_{84})$ as asymmetric error bars. We show these age measurements for the selected binnings with symmetric $\sigma_{68}$ error bars as a function of redshift in the bottom panel of Fig.~\ref{fig:d4000_age}. Note how the slopes match with the $\Lambda$CDM prediction in $0.6 < z < 0.75$, where VIPERS is magnitude complete and has most of its objects.

Working with age posteriors rather than single best-fit ages and symmetric uncertainties has several advantages in the CC context. First, it allows us to identify redshift bins where the data carry weak age information: very broad, strongly skewed, or multimodal posteriors signal that the corresponding $\Delta t$ will be poorly constrained and should be down-weighted or excluded. Second, it makes the treatment of age-metallicity degeneracy explicit: the width and asymmetry of $P(t\mid \mathrm{D})$ directly reflect how strongly D4000$_n$ and the metallicity prior constrain age in that bin. Note how in Fig.~\ref{fig:prior_posterior}, the posterior distributions for $z_{\rm mean} = 0.521$ and $0.573$ are overlapping and skewed to the right. These two bins also have the smallest number of galaxies and hence the largest $\sigma_{\rm D}$ among our $dz$ = 0.05 sample, which broadens and skews their age posteriors. The resulting statistical uncertainty in this redshift range exceeds the small age difference expected from $\Lambda$CDM over such a narrow $\Delta z$, so the apparent lack of resolved evolution here reflects insufficient signal-to-noise and not a breakdown of the underlying age-redshift relation. This flattening also occurs for the two bins on the high-redshift edge, clearly seen on Fig.~\ref{fig:d4000_age} (bottom panel). Our bin-skipping strategy in Section~\ref{SS:5.3}--\ref{SS:5.4} is specifically designed to address this. 
Furthermore, we verified that the mean stellar mass of our sample is nearly constant ($\simeq$11.17--11.22 dex) across the analyzed redshift range, with no systematic trend coincident with the observed flattening at the high- and low-redshift ends (see Appendix~\ref{Appendix:D}). It indicates a statistical rather than selection-driven origin.

\subsection{Age difference posteriors}
\label{SS:5.3}
We define the age difference between two redshift bins, $z_i$ and $z_{i+n}$ as: $\Delta t = t_{\mathrm{high}z}-t_{\mathrm{low}z}$, where $t_{\mathrm{high}z}$ and $t_{\mathrm{low}z}$ denote the ages at high and low redshifts respectively; $i$ is the bin index and $n$ is the interval between bins used to compute the age difference. When combining bins to obtain the age-difference posterior $P(\Delta t\mid \mathrm{D})$, we do not assume individual age posteriors as Gaussian. Instead, we compute $P(\Delta t \mid \mathrm{D}_i, \mathrm{D}_{i+n})$ by convolving the full posteriors $P(t \mid \mathrm{D}_i)$ and $P(t \mid \mathrm{D}_{i+n})$. This correctly propagates the non-Gaussian features (e.g., skewness) of the age posteriors into the posterior for age differences, assuming that the age posteriors in different redshift bins are statistically independent. Detailed steps to derive these $\Delta t$ posteriors at the effective redshift, $z_{\rm mid}$, between the redshift bins $z_i$ and $z_{i+n}$ are available in Appendix~\ref{Appendix:age_to_H}.

In Fig.~\ref{fig:agediff_posterior}, these posterior probability densities are shown as a function of the age difference $\Delta t$ for $dz$ = 0.05 binning. The different panels represent the cases with a different number of bins skipped ($n_{\rm skip} = n-1$), i.e., the $\Delta t$ posterior measured between $i^{\rm th}$ and $(i+n)^{\rm th}$ redshift bins. Note that we adopt the sign convention $\Delta t = t_{\mathrm{high}z} - t_{\mathrm{low}z}$. With this choice:
(i) negative values of $\Delta t$ correspond to the physical case in which galaxies at higher redshift are younger than those at lower redshift $(t_{\mathrm{high}z} < t_{\mathrm{low}z})$;
(ii) Positive values of $\Delta t$ would imply $t_{\mathrm{high}z} > t_{\mathrm{low}z}$, which is not true in standard cosmology and therefore directly quantify the posterior support for non-physical solutions. Therefore, the optimal posterior distributions are the ones that do not have a significant posterior distribution in $\Delta t \geq 0$. Keeping that in mind, the cases with $n_{\rm skip} = 1$ and $n_{\rm skip} = 2$ seem ideal, with the latter being the most ideal case.

From the age difference posteriors, we also measure the median, $\Delta t_{50}$, and the asymmetric errors, $\sigma_{\Delta t+} = \Delta t_{84} - \Delta t_{50}$ and $\sigma_{\Delta t-} = \Delta t_{50} - \Delta t_{16}$. With this, we have all the information ready to input into Eq.~\eqref{eq:cc} to calculate the $H(z)$. These $\Delta t$ values, along with their corresponding error bars, have been shown as a function of redshift in Fig.~\ref{fig:dt_vs_z} for the two selected binning strategies. The $\Lambda$CDM prediction of $\Delta t(z)$ for the corresponding measurements has also been overplotted for the ease of comparison, and it is clear that most of our measurements are consistent with the standard cosmology.

\begin{figure}
    \centering
    \begin{subfigure}{\columnwidth}
        \includegraphics[width=\columnwidth]{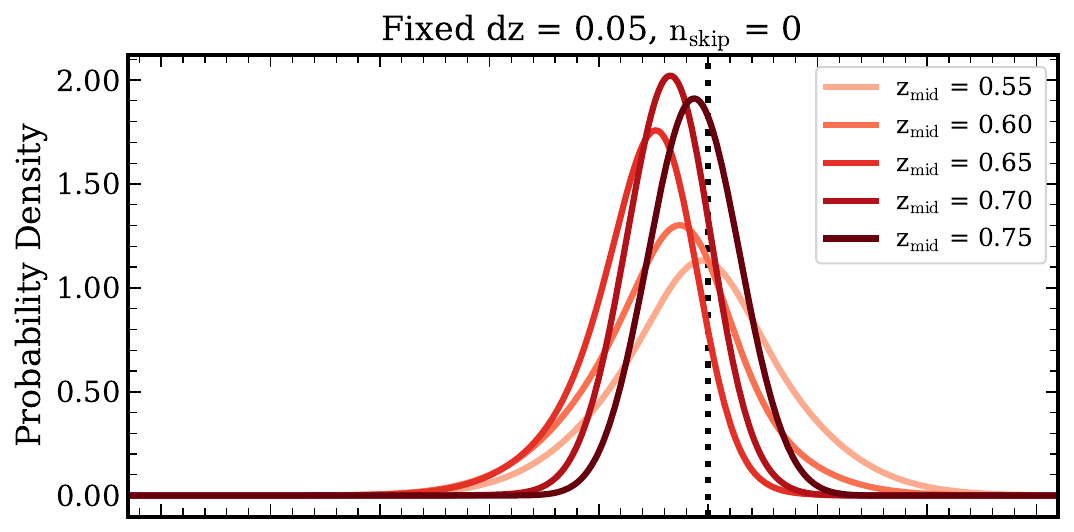}
    \end{subfigure}
    \begin{subfigure}{\columnwidth}
        \includegraphics[width=\columnwidth]{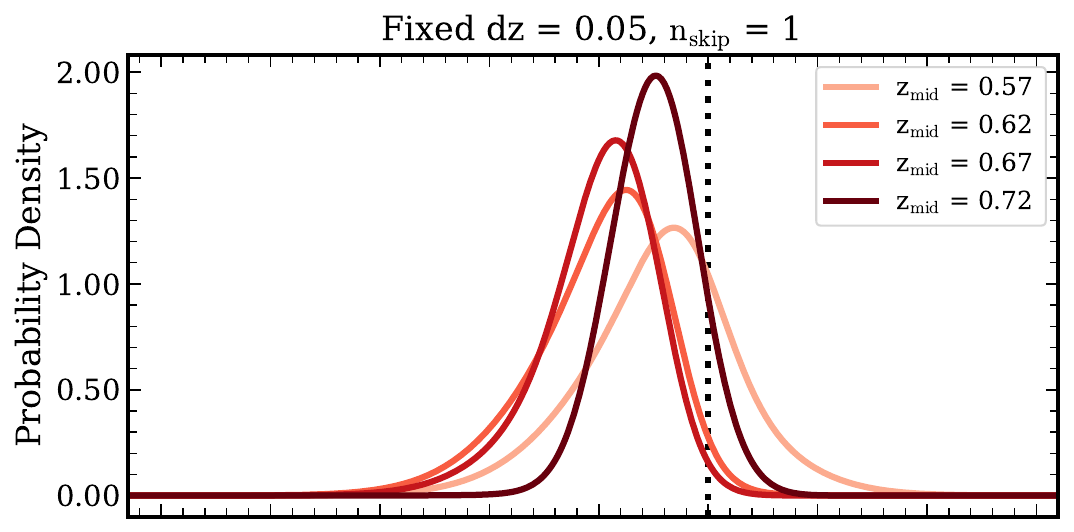}
    \end{subfigure}
    \begin{subfigure}{\columnwidth}
        \includegraphics[width=\columnwidth]{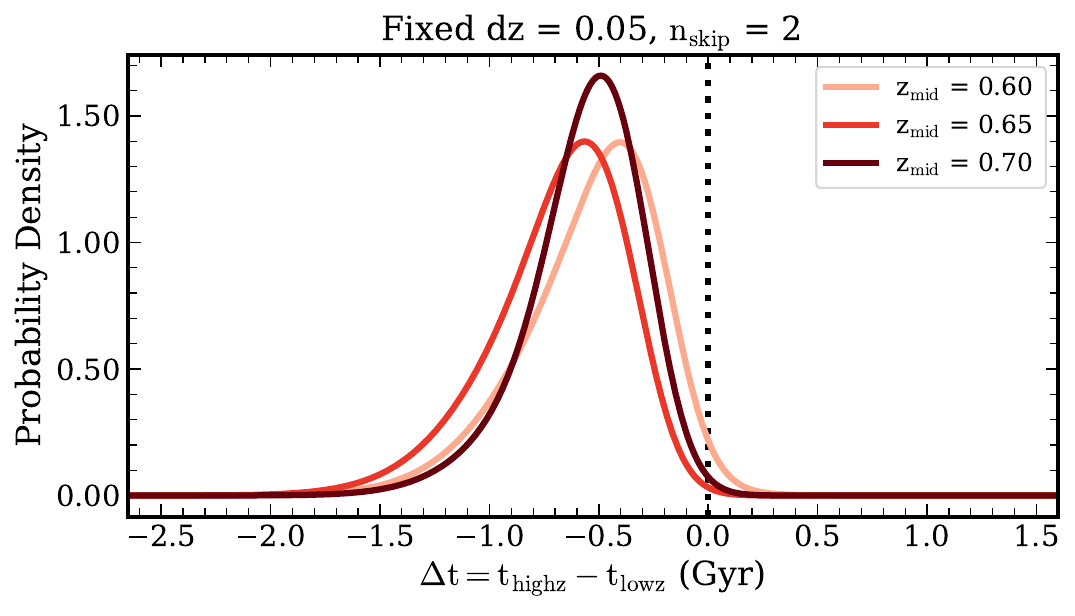}
    \end{subfigure}
    \caption{Plot of age difference posteriors for $dz$ = 0.05 binning with the top, central, and bottom panels featuring the cases with different numbers of bins skipped.}
    \label{fig:agediff_posterior}
\end{figure}

\subsection{Bin skips}
\label{SS:5.4}

Skipping bins is a very standard procedure in CC literature, and \cite{Moresco_2012b} explicitly mentions that the optimal choice of $n_{\rm skip}$ is the result of a trade-off analysis between two competing effects: we want $n_{\rm skip}$ to be as small as possible to get more number of $H(z)$ measurements, while keeping it large enough to have a significant $\mathrm{D}4000_n$ (and consequently age) evolution larger than the statistical scatter in the data. For their $z > 0.4$ sample, they found $n_{\rm skip} = 2$ to be the optimal choice. Similarly, \cite{Borghi_2022b} used $n_{\rm skip} = 1$ as their ideal choice. 

\begin{figure}
    \centering
    \begin{subfigure}{\columnwidth}
        \includegraphics[width=\columnwidth]{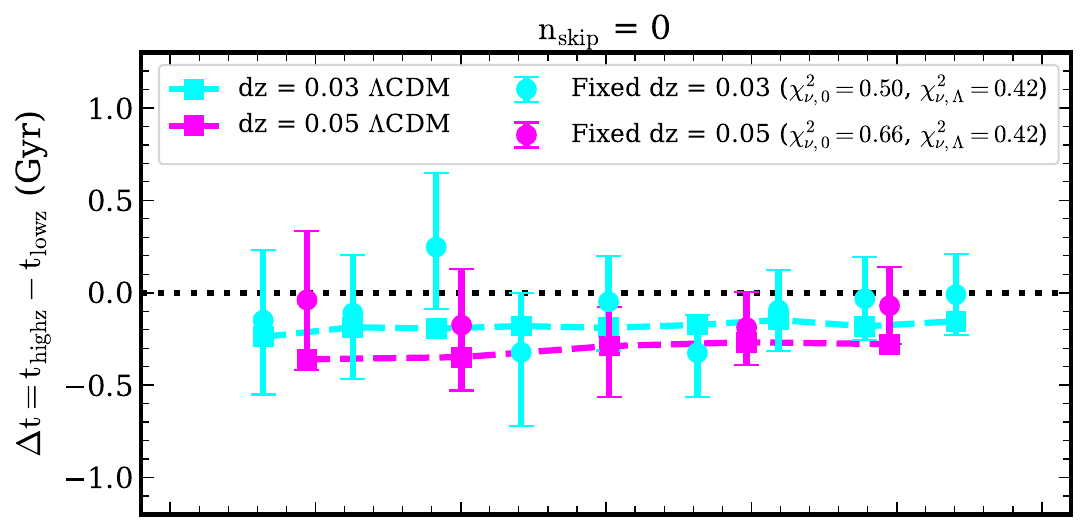}
    \end{subfigure}
    \begin{subfigure}{\columnwidth}
        \includegraphics[width=\columnwidth]{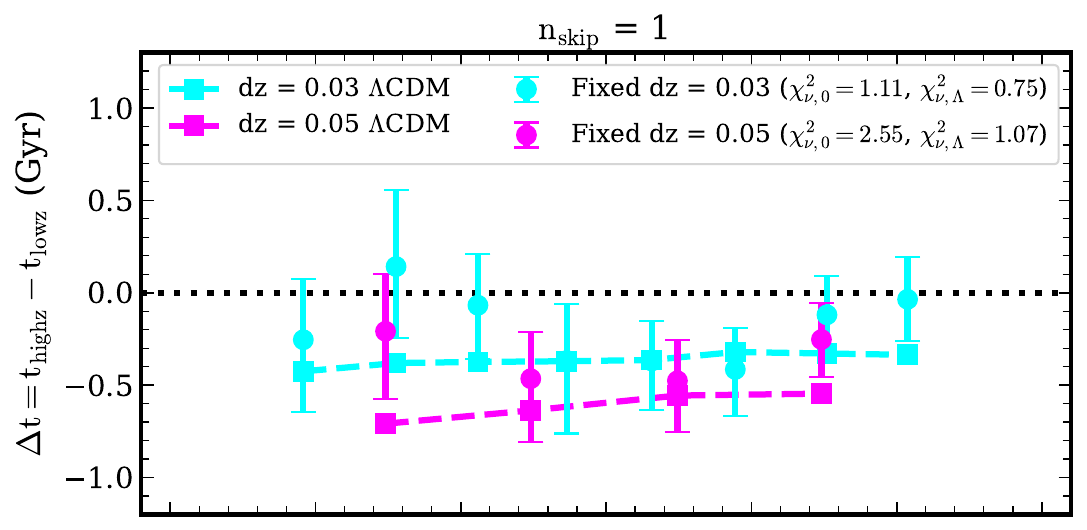}
    \end{subfigure}
    \begin{subfigure}{\columnwidth}
        \includegraphics[width=\columnwidth]{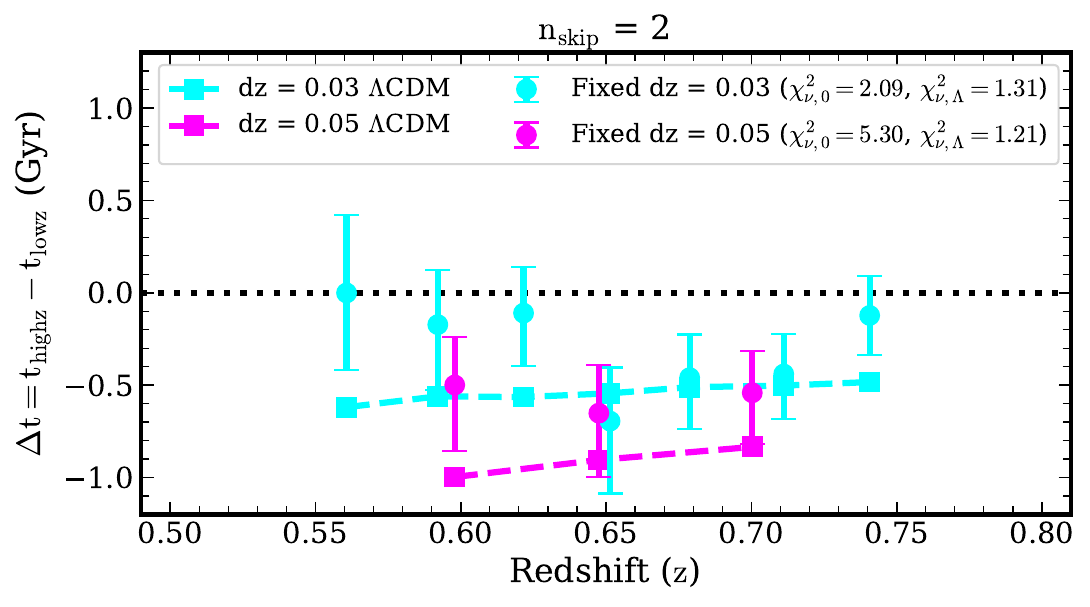}
    \end{subfigure}
    \caption{Plot of comparison of $\Delta t$ vs. $z$ for three different bin skip steps ($dz$ = 0.05 binning) with the top, central, and bottom panels featuring the cases with different numbers of bins skipped.}
    \label{fig:dt_vs_z}
\end{figure}

To evaluate different bin skips ($n_{\rm skip}$), we use a specific metric: the reduced $\chi^2$ fit to both $\Delta t=0$ ($\chi^2_{\nu, 0}$) and the $\Lambda$CDM prediction ($\chi^2_{\nu, \Lambda}$). The $\chi^2$ values are displayed, together with both predictions, in Fig.~\ref{fig:dt_vs_z}. 
Note that the measurements with $\chi^2_\nu$ values $\sim 1$ are more consistent, and those farther from 1 are least consistent. As per this principle, the $n_{\rm skip} = 0$ case is the least compatible one since the measurements are more consistent with $\Delta t=0$ (non-expanding universe) than with standard cosmology. For $n_{\rm skip} = 1$, $dz$ = 0.03 binning is not optimal since it aligns well with $\Lambda$CDM but also shows strong consistency with $\Delta t = 0$. In contrast, $dz$ = 0.05 is very optimal, demonstrating high $\Lambda$CDM consistency while being less compatible with a non-expanding universe. In case of $n_{\rm skip} = 2$, both $dz$ = 0.03 and $dz$ = 0.05 binnings are highly consistent with $\Lambda$CDM compared to $\Delta t = 0$.

We highlight that the goal of this analysis is not to choose a priori the binning+$n_{\rm skip}$ combination that better fits $\Lambda$CDM, but to evaluate the ones that can properly constrain the expansion of the Universe. Combinations of $n_{\rm skip}$ and binning compatible with Delta $t = 0$ are less likely to detect any significant expansion, as they are compatible with a model where time evolution is decoupled from redshift. Combinations compatible with $\Lambda$CDM will also be compatible with any other model of expansion within our precision limits. Therefore, we keep all binning+$n_{\rm skip}$ combinations until the end of our analysis.

\section{Hubble rate estimation}
\label{S:6}

\subsection{Error budget}
\label{SS:6.1}

Before we proceed to derive the Hubble rate, it is important to quantify all the uncertainties involved in our calculations. There are two main components of these: (i) statistical errors arising from the noise in the data (e.g., uncertainty in the measurement of D4000$_n$, which is propagated from the pixel-level error in flux/SED); and (ii) systematic errors, which are consistent inaccuracies that can not be removed just by adding more objects to the dataset (e.g., in the context of galaxy evolution these are usually calibration errors that can arise from the specific SSP model characteristics and their parameters (IMF, SFH, etc.), stellar libraries, and degeneracies.). 

In CC studies, the statistical uncertainties ($\sigma_{\rm stat}$) arise directly from the D4000 measurement errors and galaxy sample variance, which depend on the characteristics of the survey dataset being used. This error finally appears in the CC equation (Eq.~\ref{eq:cc}) as the statistical error in the measurement of age difference, $dt$, and is propagated into the total error, $\sigma_{\rm H} = \sigma_{\rm stat} + \sigma_{\rm syst}$. The other component ($\sigma_{\rm syst}$), which comes from SSP models and the metallicity component, has been extensively studied in the CC literature and explicitly quantified by \cite{Moresco_2020} after examining multiple models and their parameters. In existing studies, the metallicity error is usually incorporated into the systematics, whereas our marginalization framework accounts for it while calculating ages. Therefore, the metallicity contribution is already intrinsic to our $dt$ measurement error, but we still need to incorporate the systematic errors independently. 

One recent example where systematic errors were calculated using a covariance matrix structure as proposed by \cite{Moresco_2020} is the study by \cite{Loubser_2025b}, which calculated systematic errors as: $\sigma_{\rm syst} = 3\%\, (\text{SFH}) \pm 7\%\, (\text{library}) \pm 6\%\, (\text{model}) \pm 10\%\, (\text{metallicity})$. These errors were then added in quadrature to get the total systematic error, which is again added in quadrature to the statistical error to get the total error on $H(z)$. Since all our D4000$_n$ measurements are well within the range tested by \cite{Moresco_2020}, we adapt their suggested systematic error contributions and get (average errors in $0 \le z \le 1.5$): $\sigma_{\rm syst} = 0.4\%\, (\text{IMF}) \pm 2.5\%\, (\text{SFH}) \pm 6.5\%\, (\text{st. library}) \pm 9\%\, (\text{SPS model})$. Summing them in quadrature amounts to a total systematic error of $\sim 11.39\%$. 

Unlike previous CC studies that fix metallicity to a single value and add a corresponding empirical systematic term (e.g., \citealt{Moresco_2012b}, \citealt{Loubser_2025b}), our framework marginalizes the full joint posterior $P(t, Z\mid D)$ over metallicity using an explicit Gaussian prior (Section~\ref{SS:5.2}). The resulting metallicity uncertainty is therefore already propagated into the width and shape of our age posteriors, and hence into our quoted statistical error on $H(z)$. Adding a further metallicity systematic term in quadrature on top of this would double-count the same source of uncertainty.

We further emphasize that in the tests carried out with modified priors (Appendix~\ref{Appendix:C}; mean shifted, width doubled) we deliberately use extreme prior perturbations (e.g., a $0.05$~dex mean shift corresponds to $\sim1.56\sigma$ relative to our fiducial uncertainty) to verify whether our results are still robust. Even under the most extreme case, $H(z)$ remains statistically consistent with both our fiducial result and $\Lambda$CDM (within $1\sigma$ and $1.2\sigma$, respectively). We do not add this as a separate quadrature systematic term, since doing so would also exaggerate the error budget by unnecessarily accounting for variations beyond our reasonable choice of prior.

\subsection{Results}
\label{SS:6.2}

With all the ingredients in place, we now follow the procedure outlined in Appendix~\ref{Appendix:age_to_H} to derive $H(z)$ and its statistical uncertainty from the age-difference posteriors $P(\Delta t\mid \mathrm{D})$ for each pair where age difference is computed (as in Fig.~\ref{fig:agediff_posterior}). The redshift of each one of these individual H(z) measurements is the median redshift of the two bins used to compute $P(\Delta t\mid \mathrm{D})$. We carry out this analysis for each binning+$n_{\rm skip}$ combination considered in the previous section. This results in a total of 6 sets (one set for each binning+$n_{\rm skip}$ combination) of $H(z)$ estimates with their respective errors.

\begin{figure}
    \includegraphics[width=\columnwidth]{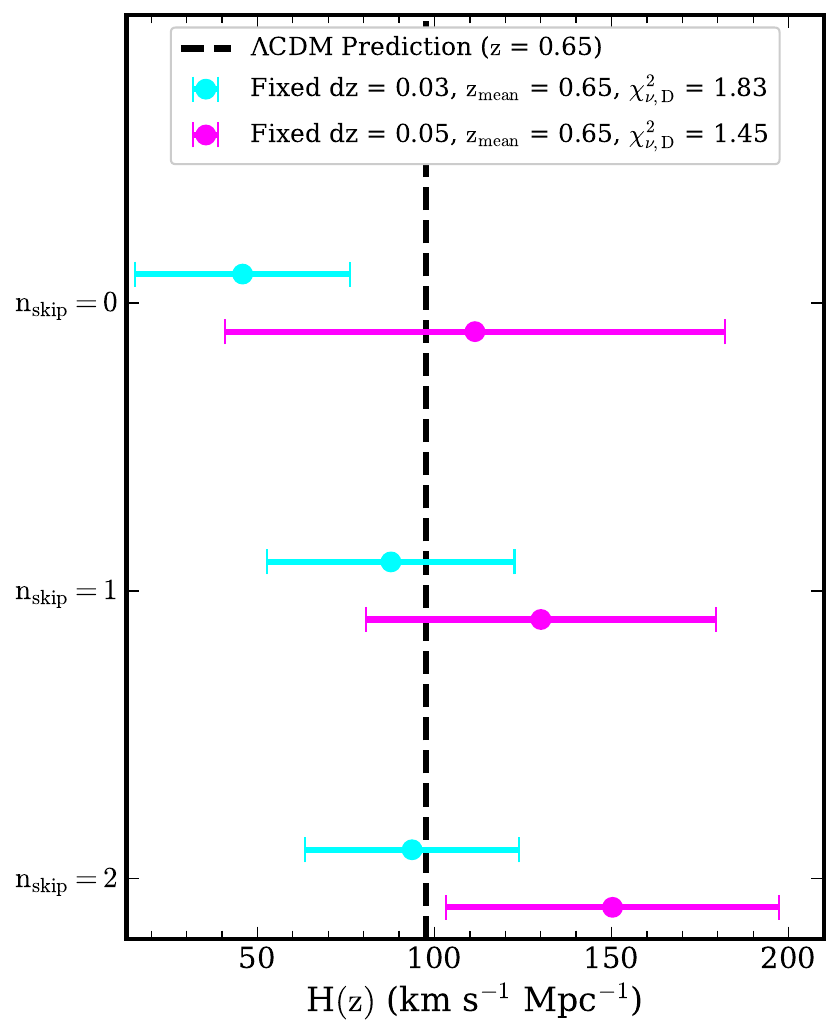}
    \caption{Plot of weighted average H($z$) for the two best binning schemes (from Section~\ref{SS:5.1}) and different bin skip steps. The error bars represent the symmetric $\sigma_{68}$ statistical errors combined in quadrature with systematic errors.}
    \label{fig:wavg_H}
\end{figure}

Because different bin-pair combinations probe overlapping or complementary redshift ranges, we combine these individual estimates into a single robust $H(z)$ measurement via an inverse-variance weighted average,
\begin{equation}\label{eq:ivwa}
    \bar{H}(z) = \frac{\sum_i w_i\, H_i(z)}
                      {\sum_i w_i}\,,
    \qquad w_i = \frac{1}{\sigma_{\mathrm{H},i}^{2}}\,,
\end{equation}
where $\sigma_{H,i}$ is the statistical uncertainty on the $i^{\rm th}$ estimate. The corresponding uncertainty on the weighted average is
\begin{equation}\label{eq:ivwa_err}
    \sigma_{\bar{\rm H}}^{2} = \frac{1}{\displaystyle\sum_i w_i}\,.
\end{equation}
This weighting scheme suppresses the contribution of poorly constrained estimates (those with large $\sigma_{\mathrm{H},i}$, typically arising from bins with low galaxy counts or broad age posteriors) while preserving the full statistical weight of well-constrained measurements. The result is a single, statistically robust $H(z)$ determination at an effective redshift, which we can compare directly with independent CC measurements from the literature and with other probes of $H_0$. We do this for the two fixed-width binning strategies identified as optimal in Section~\ref{SS:5.1} ($dz$ = 0.03 and $dz$ = 0.05). The results are shown in Fig.~\ref{fig:wavg_H}. 

Several features are immediately apparent. First, all six measurements are consistent with the Planck $\Lambda$CDM prediction at an effective redshift, $z_{\rm eff} = 0.65$, within $\lesssim 1.2\sigma$, confirming the overall compatibility of our D4000-based chronometer analysis with the standard cosmological model. Second, within each binning strategy, the three $n_{\rm skip}$ values yield mutually consistent $H(z)$ estimates, proving that our methodology is self-consistent.

However, not all binning+$n_{skip}$ choices are optimal. If the redshift differential $\Delta z$ is too small (either because of too narrow redshift bins and/or too few skips), the absolute value of $\Delta t$ is too small compared to the statistical error (i.e., too low S/N), and may be consistent with $\Delta t = 0$, as discussed in Section~\ref{S:5}. Since the error of $H(z)$ is proportional to $\Delta t^{-1}$ (see Appendix~\ref{Appendix:age_to_H}), if the S/N of $\Delta t$ is too low, the bins with $\Delta t$ closer to zero will have disproportionately large $H(z)$ error. These will also be the bins with greatly overestimated $H(z)$ (Eq.~\ref{eq:cc}), and since they will have negligible weights when stacking via inverse-variance weighting (Eq.~\ref{eq:ivwa}), the stacked $H(z)$ will be biased toward lower values. We see this trend in Fig.~\ref{fig:wavg_H} for $dz=0.03$ and $n_{skip}=0$, and we found that this negative bias is even more pronounced for narrower bins.
 
On the other hand, if $\Delta z$ is too large (due to too wide bins or too many bins skipped), the assumptions behind CCs may not fully hold, introducing biases in the $H(z)$ measurement. For example, the progenitor bias \citep[see][]{Moresco_2012b}: if $\Delta z$ is too large, the younger galaxies in the low-$z$ bin may not have progenitor counterparts in the high-$z$ bin. This lack of youngest progenitors will result in an underestimation of $\Delta t$ (as in the high-$z$ bin, the median galaxy age will be older than expected), and thus an overestimation of $H(z)$. We can observe this trend in Fig.~\ref{fig:wavg_H}, where the values of $H(z)$ systematically increase with the number of skips (especially for $dz=0.05$).

With these caveats in mind, we consider that the best measurement in Fig.~\ref{fig:wavg_H} is the one for fixed $dz$ = 0.03 with $n_{\rm skip} = 2$, which has the smallest error bar, and the mean value is highly consistent with $\Lambda$CDM. This measurement at $z = 0.65$ we have is
\begin{equation}
    H(z=0.65) = 93.68 \pm 28.27\ \pm 10.67\, \mathrm{km\,s^{-1}\,Mpc^{-1}}.
\end{equation}

We note that the uncertainties shown in Fig.~\ref{fig:wavg_H} are a combination of statistical uncertainties propagated from the age-difference posteriors $P(\Delta t\mid \mathrm{D})$ as described in Appendix~\ref{Appendix:age_to_H}, and the systematic uncertainties. The systematic error budget, quantified via the covariance matrix proposed by \cite{Moresco_2020}, is 11.39\%, which corresponds to an absolute systematic error of 
\begin{equation}
    \sigma_{\rm syst} = 0.1139 \times 93.68
    \approx 10.67\ \mathrm{km\,s^{-1}\,Mpc^{-1}}.
\end{equation}
Assuming the statistical and systematic errors are independent and
approximately Gaussian, the total uncertainty is obtained by adding them
in quadrature:
\begin{equation}
    \sigma_{\rm H}
    = \sqrt{\sigma_{\rm stat}^{2} + \sigma_{\rm syst}^{2}}
    \approx \sqrt{28.27^{2} + 10.67^{2}}
    \approx 30.22\ \mathrm{km\,s^{-1}\,Mpc^{-1}}.
\end{equation}

The total error budget is therefore dominated by the statistical component, and the plot presented in Fig.~\ref{fig:wavg_H} is representative of our final constraints. We report both the statistical and systematic errors for our best measurement as our final measurement constraint in Table~\ref{tab:Hz_data}.

\begin{figure}
\includegraphics[clip,width=\columnwidth]{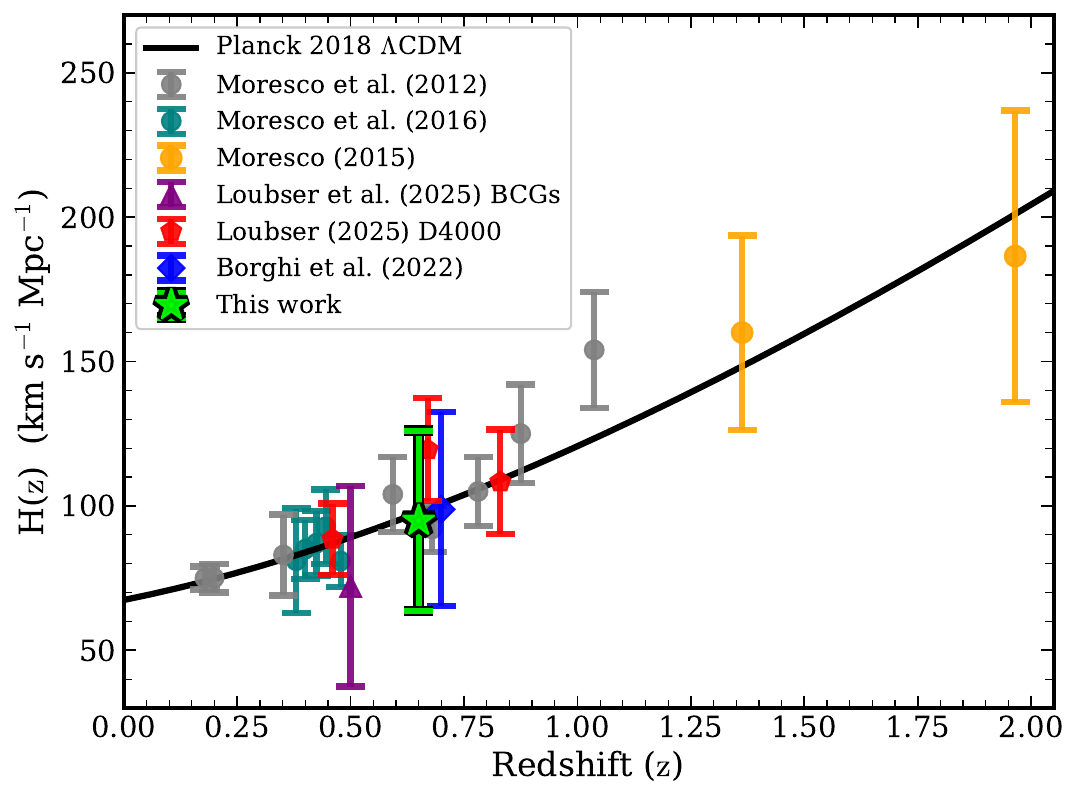} 
\caption{The $H(z)$ measurement from our work compared with various cosmic chronometer studies. The values plotted are presented in Table~\ref{tab:Hz_data}.}
\label{fig:cc_fin}
\end{figure}

Figure~\ref{fig:cc_fin} places our result in the context of existing CC measurements spanning $0 \lesssim z \lesssim 2$, alongside the Planck 2018 $\Lambda$CDM prediction. Our measurement at $z_{\rm eff} \simeq 0.65$ is fully consistent with the $\Lambda$CDM curve to within $\lesssim 1\sigma$ and occupies a region of intermediate redshift space that is already probed by the spectroscopic studies of \cite{Moresco_2011, Moresco_2012b, Moresco_2015, Moresco_2016} and the D4000-based constraint of \cite{Loubser_2025b}, as well as the result of \cite{Borghi_2022b} from the Lick indices method and full-spectral fitting done by \cite{Jiao_2023}. Our statistical error bar is comparable in size to those of the existing measurements at similar redshifts, demonstrating that the D4000 CC method applied to the VIPERS passive galaxy sample achieves a competitive precision for a single-bin photometric-quality measurement. Encouragingly, all measurements shown are mutually consistent within their respective uncertainties, reinforcing the overall picture of a smoothly evolving expansion history in broad agreement with the standard cosmological model. 
The present constraint therefore provides independent validation of the existing CC compilation at $z \simeq 0.65$ and highlights the possibility that low- to medium-resolution spectroscopy and photometric-only samples can contribute meaningful CC data points as a complement to the deep, high-resolution spectroscopic measurements that have defined the field to date.

\section{Discussion}
\label{S:7}

\subsection{New developments and advances}

This work presents one of the first applications of the cosmic chronometers method to a moderate-resolution spectroscopic survey sample using only color and mass cuts for sample selection that can be reproduced with photometric surveys. For photometric surveys such as PAUS, reasonably accurate photometric redshifts are available, which have been used for determining broadband magnitudes and stellar masses \citep{Eriksen_2019, Renard_2022_D4000, Navarro_2024}. The first study using a photometric sample for CC was by \cite{Jimenez_2023}, which applied a neural network trained on a spectroscopic sample to the \texttt{COSMOS2015} survey to create an age-redshift relation. Our approach, on the other hand, is more direct and a simple modification of the standard D4000 approach. Previous studies have relied almost exclusively on high-resolution deep spectroscopic datasets \citep{Moresco_2012b, Moresco_2016, Borghi_2022b, Loubser_2025b}, which are observationally expensive. In contrast, we use VIPERS PDR2, which has low- to medium-resolution spectroscopy using only the rest-frame colors, D4000$_n$ values, and spectroscopic redshifts; all of these quantities can also be obtained to reasonable accuracies with a purely photometric dataset (narrow-band photometry for D4000, e.g., \citealt{Renard_2022_D4000}). Therefore, this work is not only an independent spectroscopic measurement of $H(z)$ with a novel Bayesian framework, but also establishes the basis for a photometry-only measurement in our upcoming work.

Another novelty of this work is the way we treat age uncertainties. Rather than adopting single best-fit ages with Gaussian error propagation, we propagate the full, non-Gaussian age posteriors $P(t_i\mid \mathrm{D}_i)$ into the age-difference posterior $P(\Delta t\mid \mathrm{D})$ via a direct numerical convolution (Appendix~\ref{Appendix:D_to_age}), and report the resulting $H(z)$ constraints at the median and 16$^{\rm th}$/84$^{\rm th}$ percentile levels ($\sigma_{68}$). This approach captures asymmetric uncertainties arising from the D4000-age relation and the age-metallicity degeneracy in a self-consistent manner, a feature largely absent from earlier CC analyses based on D4000.

\subsection{Advantages and limitations}

\subsubsection{Advantages}
\label{SSS:Advantages}

\begin{itemize}
      \item \textit{Uses D4000 as the main observable.} This study uses D4000 for massive, passive galaxies, preserving the main robustness advantage of the standard D4000 CC method over full spectral fitting (simple, high-S/N observable, weak sensitivity to flux calibration and dust in the CC regime)\footnote{See, e.g., \citet{Moresco_2012a, Moresco_2015}.}. Moreover, D4000 could also be measured with sufficient precision using narrow-band photometry alone.

      \item \textit{Bayesian treatment of D4000 and metallicity uncertainties.} Constructs full posteriors $P(t,Z\mid \mathrm{D})$ and marginalized $P(t\mid \mathrm{D})$, naturally capturing non-Gaussian and asymmetric age uncertainties, rather than enforcing a linear D4000-age relation with Gaussian errors.
      
      \item \textit{Fully convolved age difference posteriors.} Derives the distribution of $\Delta t$ as the cross-correlation of age posteriors in neighboring redshift bins, ensuring that the $\Delta t$ uncertainty (and its potential skewness) is correctly propagated into $H(z)$ rather than approximated by simple variance addition. 
      
      \item \textit{Cosmology independent.} Keeps the CC method cosmology independent by restricting priors to purely astrophysical constraints (e.g., non-negative ages, broad age ranges, metallicity priors from independent spectroscopy) and avoiding any prior that depends on $(\Omega_{m},\Omega_\Lambda,w(z))$ or on a model-dependent age-redshift relation.
      
      \item \textit{Highly modular framework.} Provides a clean modular structure (observables, SPS forward model, priors, derivation of $H(z)$ from kinematics of expansion), making it straightforward to swap SPS grids, metallicity priors, or binning schemes and to quantify their impact on the final $H(z)$ constraints.
      
\end{itemize}

Similar to the full spectral fitting method for CC, our framework uses a joint posterior over age and metallicity (Eq.~\ref{eq:joint_posterior}), rather than assuming a single fixed metallicity as in direct D4000-differencing approaches. However, unlike full spectral fitting, we constrain metallicity via an external prior rather than fit self-consistently alongside age from the full spectrum. This is a necessity rather than a choice since VIPERS's spectral resolution ($R\approx220$) cannot support reliable estimation of per-galaxy metallicity-sensitive indices, and the available rest-frame photometric SED fitting \citep{Moutard_2016a, Moutard_2016b} constrains stellar mass and luminosity but not metallicity for our passive sample. A central motivation of this work is precisely to sidestep full spectral fitting, since it is only applicable to high-resolution spectroscopic surveys and therefore restricts the cosmic chronometer method to a narrow class of expensive, small-volume spectroscopic datasets. In contrast, our hybrid scheme is designed to extend the method to moderate- and low-resolution spectroscopic and photometric surveys. We therefore adopt an external prior calibrated on a massive, passive population \citep{Citro_2016}. In principle, this also extends to metallicity the same robustness against flux-calibration and extinction systematics that D4000 already provides for age, rather than inheriting these systematics through a joint fit. When compared to the standard direct D4000-differencing approach \citep[e.g.,][]{Moresco_2011, Moresco_2012a, Moresco_2012b, Moresco_2013}, our full Bayesian treatment avoids linearizing the D4000--age relation (which saturates at old ages. Fig.~\ref{fig:d4000_age}), correctly propagates the non-Gaussian age uncertainties seen in Fig.~\ref{fig:prior_posterior}, and convolves full age posteriors (Eq.~B.15) rather than differencing point estimates, so that posterior skewness is preserved in $\Delta t$ and $H(z)$.

\subsubsection{Remaining limitations}
\label{SSS:Limitations}

\begin{itemize}
      \item \textit{Young stellar component contamination.} Residual contamination from a young stellar component ($\lesssim 2\, \mathrm{Gyr}$) in nominally passive galaxies may bias D4000$_n$ toward lower values and hence biases ages toward lower values. Our selection criteria (Section~\ref{S:3}) are designed to minimize this. While a fraction of the sample would not be considered passive enough for CCs with spectroscopic criteria, this contamination does not affect our results (as proven when repeating the analysis with a much tighter spectroscopic cut in Appendix~\ref{Appendix:spec_sel}). However, we cannot rule out that this contamination may become a problem for tighter $H(z)$ constraints (e.g., with larger samples).

      \item \textit{Single SPS model.} Still relies on a specific SPS grid $D_{\mathrm{model}}(t,Z)$; systematic differences between SPS models, stellar libraries, and IMFs remain one of the main sources of uncertainty and are not automatically removed by the Bayesian treatment (they could be explored or marginalized over different stellar libries and SPS models and their parameters in upcoming work). Tentatively, the systematic error contributions have been adapted from \cite{Moresco_2020}.
      
      \item \textit{Binned D4000.} Uses binned, median D4000 measurements per redshift bin, which discards information in the full galaxy-by-galaxy D4000 distribution and may hide residual trends with mass, color, or subtle contamination by young components within a bin. However, this is a standard practice in all CC studies, which also takes the median D4000 and MAD errors. They do this because the median is much less sensitive to such outliers than the mean, so it better represents the ``typical'' passive ETGs in that bin. The distribution of D4000 values within a bin is often skewed or has extended tails (e.g., because of residual SF or metallicity spread), so using the median and the median absolute deviation (MAD) gives an uncertainty estimate that is robust to non-Gaussian scatter, rather than relying on the standard deviation, which is very sensitive to tails. Similarly, we do not want error bars dominated by a small number of problematic galaxies and adopt the same approach.
    
      \item \textit{Requires priors.} Since low- to medium-resolution spectroscopy and photometry do not allow for direct metallicity measurements, our method requires age and metallicity priors. If the observational data are highly informative (high-S/N D4000, or a very large and well-curated sample), then these priors impose an upper bound in the precision of our $H(z)$ constraints, so they must be justified and tested for robustness. In our case, the age prior (top-hat or flat prior) is justified, since it only selects an age range (e.g., 0 to 14 $\mathrm{Gyr}$) to rule out impossible age values, but does not bias toward any specific age value. The metallicity prior can be justified by shifting the mean (e.g., by 0.05 dex), widening the error (e.g., by a factor of 2), and showing that the results remain consistent (see Appendix~\ref{Appendix:C}).
      
\end{itemize}

\subsection{Comparison with independent cosmological probes}
Our $H(z)$ measurement ($H(z = 0.65) = 93.68\pm28.27\pm10.67\, \mathrm{km\,s^{-1}\,Mpc^{-1}}$) can be translated into a constraint on the matter density $\Omega_{m}$ (assuming a flat $\Lambda$CDM cosmology and fixed $H_0$), or used directly as a data point in cosmological fits. While a full cosmological parameter analysis is beyond the scope of this paper, we note that our result is broadly consistent with the parameter constraints inferred from the following independent probes. 

\textit{Cosmic microwave background.}
The Planck 2018 TT+TE+EE+lowE likelihood, combined with lensing and BAO, yields $H_0 = 67.4 \pm 0.5$\,km\,s$^{-1}$\,Mpc$^{-1}$ and $\Omega_{m} = 0.315 \pm 0.007$ within a flat $\Lambda$CDM model \citep{Planck_2020_2018_results}, implying $H(0.65) \simeq 97.7$\,km\,s$^{-1}$\,Mpc$^{-1}$, with which our result agrees at $\lesssim 1\sigma$. The CMB measurement probes the expansion history at the surface of last scattering ($z \simeq 1100$) and constrains $H(z)$ at lower redshifts only through model-dependent extrapolation, making our direct CC measurement at $z \sim 0.65$ a valuable independent cross-check.

\textit{Baryon acoustic oscillation.} 
BAO measurements constrain the combination $H(z)\, r_{\rm d}$, where $r_{\rm d}$ is the sound horizon at the drag epoch. BOSS DR12 \citep{Alam_2017_BAO} measured $H(z) r_{\rm d}$ to $\sim$2--3\% precision over $0.2 < z < 0.75$, and the first-year DESI results \citep{Adame_2025_DESI} have extended this to $z \sim 4$ with comparable or better precision. When combined with a sound-horizon calibration from the CMB, these translate to $H(z)$ values fully consistent with $\Lambda$CDM and with our measurement at $z \simeq 0.65$. Crucially, CC constraints require no assumption about $r_{\rm d}$ or any early universe physics, making them complementary to BAO in cosmological fits \citep{Moresco_2022}.

\textit{Type Ia supernovae.}
The Pantheon+ compilation \citep{Scolnic_2022_SNI} of 1550 spectroscopically confirmed SNe\,Ia constrains the luminosity distance over $0.001 < z < 2.26$; the Union3 compilation \citep{Rubin_2025_SNI} provides a complementary dataset of 2087 SNe\,Ia. 
However, these measurements are only relative distances (i.e.,\ $H(z)/H_0$) and an external $H_0$ anchor is required for calculating $H(z)$. When anchored to BAO+CMB measurements from \cite{Planck_2020_2018_results} ($H_0 = 67.4\pm0.5$ km s$^{-1}$ Mpc$^{-1}$), SNeIa measurements are consistent with the early universe measurements (e.g., BAO, CMB). In contrast, anchoring to late-universe distance ladder measurements, such as \texttt{SH0ES} \citep{Riess_2022_SNI}, renders them incompatible with early universe measurements, resulting in the Hubble tension. Cosmic chronometer methods have not yet reached the precision required to resolve this tension, and the preference for either early or late-universe measurements has not been well established, but they provide a cosmology-independent picture.

Beyond comparisons with other cosmological probes, our result can also be placed alongside recent CC measurements that extend the method to new tracer populations. \citet{Tomasetti_2026} applied the CC method to massive galaxy clusters, obtaining $H(z=0.542) = 66^{+81}_{-29} (\text{stat.}) \pm 13 (\text{syst.}) \mathrm{~km\ s^{-1}\ Mpc^{-1}}$. In another study, \citet{Lisiecki_2026} demonstrated how compact relic galaxies can act as chronometers, constraining $H(z=0.15) = 85\pm53 \mathrm{~km\ s^{-1}\ Mpc^{-1}}$. Both measurements are statistically consistent with our own result and with $\Lambda$CDM, reinforcing the emerging picture of compatible $H(z)$ constraints from diverse chronometer populations at low-to-intermediate redshift.
Our measurement therefore joins a consistent picture from multiple independent probes at intermediate redshifts, providing a further model-independent constraint on the expansion history at $z \simeq 0.65$.

\subsection{Applicability to photometric surveys}

The framework developed in this work is naturally extendable to future spectro-photometric and narrow-band photometric surveys. In particular, surveys such as PAUS \citep{castanderPAUCameraPAU2012, Serrano_2023, Navarro_2024, castanderPAUSurveyPhotometric2024} and J-PAS \citep{Benitez_2014_JPAS, taylorJpcam12Gpixel2014} are especially well suited for this approach because their filter sets provide sufficient spectral resolution to measure the 4000\,\AA\ break directly from photometry. PAUS has already demonstrated that D4000 can be recovered from narrow-band photometry and used to study the properties of passive galaxies in combination with spectroscopic calibration data \citep{Renard_2022_D4000}. Likewise, J-PAS will survey a much larger area with 54 narrow bands and very accurate photometric redshifts, making it a powerful dataset for identifying passive galaxies and extending D4000-based chronometer analyses to larger samples and potentially a broader redshift range \citep{Benitez_2014_JPAS}. Moreover, other deeper ongoing narrow-band surveys, such as ODIN \citep{Lee_2024_ODIN} (with 3 narrow bands) and Hyper Suprime-Cam Subaru Strategic Program \citep[HSC-SSP;][]{Aihara_2022_Subaru} with wider bands, might be able to measure D4000 to the required precision for CC over a wider redshift range.

These surveys offer an important opportunity to move the cosmic chronometer method beyond purely spectroscopic samples. In this context, our Bayesian framework is particularly valuable because it propagates non-Gaussian age posteriors, marginalizes over metallicity, and can be applied consistently to large photometric datasets, where selecting passive galaxies and calibrating D4000 remain the dominant challenges. With improved control of stellar population systematics, future spectro-photometric surveys may provide CC constraints with much smaller statistical errors and greatly expanded sky coverage. We will also apply our framework to PAUS (narrow-band photometry) in our upcoming work.

\section{Conclusion}
\label{S:8}

In this work, we have presented a new measurement of the Hubble parameter $H(z)$ at $z_{\rm eff}\simeq0.65$ using the cosmic chronometer method applied to passive galaxies from the VIPERS PDR2 survey. Most existing CC studies rely on high-resolution spectroscopy to accurately distinguish between passive and star-forming galaxies, obtain precise D4000 measurements, and constrain stellar metallicities to mitigate the age-metallicity degeneracy.  
The central question explored here is whether competitive CC constraints can still be obtained from surveys that do not provide high-quality spectroscopy, but nevertheless deliver sufficiently accurate redshifts and D4000 measurements, potentially using photometric observables alone.

Starting from the VIPERS PDR2 catalog, we selected a sample of massive, passively evolving galaxies over $0.5 \le z \le 0.8$ using color--color and stellar-mass criteria designed to minimize contamination from star-forming or recently quenched systems (Section~\ref{S:3}).  
We then measured median D4000$_n$ values in a series of fixed-width redshift bins spanning $\Delta z = 0.01$--$0.09$, and evaluated the consistency of different binning schemes against the Planck 2018 $\Lambda$CDM prediction using the reduced $\chi^2_\nu$ statistic (Section~\ref{S:4}).

The core of our analysis is a Bayesian inference framework in which the observed D4000$_n$ values are compared against \texttt{BC03} SPS model grids in the age-metallicity plane.  
Rather than fixing the metallicity of the CC population, we marginalized over it using a Gaussian prior on $\log(Z/Z_\odot)$ motivated by the measurements of \cite{Citro_2016} for massive passive VIPERS galaxies.  
The resulting age posteriors, $P(t_i|\mathrm{D}_i)$, are generally non-Gaussian and are propagated into age-difference posteriors, $P(\Delta t|\mathrm{D})$, via direct numerical convolution without assuming Gaussianity or correlations between neighboring bins (Section~\ref{SS:5.2}).

From these age-difference posteriors, we derived $H(z)$ estimates for different combinations of redshift bins characterized by a bin-skip parameter $n_{\rm skip}=0,1,2$, and combined them into a final inverse-variance-weighted measurement (Section~\ref{S:6}).  
Our main conclusions are summarized below.

\begin{enumerate}

\item \textit{Consistency with $\Lambda$CDM.}

For our preferred binning strategies ($\Delta z = 0.03$ and $\Delta z = 0.05$), and for all three bin-skip values, the inferred $H(z)$ measurements are consistent with the Planck 2018 $\Lambda$CDM prediction at $z\simeq0.65$ within $\lesssim1.2\sigma$.  
This provides an additional model-independent constraint on the expansion history at intermediate redshifts.

\item \textit{Competitive statistical precision.}

Our final measurement is comparable in precision to existing spectroscopic CC constraints at similar redshifts \citep{Moresco_2012b, Moresco_2016, Borghi_2022b, Loubser_2025b}.  
This demonstrates that photometric-quality D4000$_n$ measurements from intermediate-resolution surveys such as VIPERS can already yield competitive CC constraints without requiring the full spectroscopic depth and resolution of dedicated large observational programs such as DESI \citep{DESI_2016}.

\item \textit{Robustness to metallicity assumptions.}

Varying the metallicity prior mean by $\pm0.05$ dex or doubling its width changes the inferred $H(z)$ value by at most $\sim26$\%, corresponding to less than $1.2\sigma$ in statistical significance (see Appendix~\ref{Appendix:C}).  
This indicates that the inference is primarily driven by the data likelihood rather than by the adopted metallicity prior.

\item \textit{Systematic uncertainty budget.}

The total uncertainty remains dominated by statistical errors.  
The systematic component, estimated using the covariance-matrix formalism of \cite{Moresco_2020}, includes contributions from SPS model choice, stellar libraries, and the IMF.  
No separate metallicity systematic is added because metallicity uncertainties are already marginalized within the Bayesian framework itself. 
Future improvements in SPS modeling will therefore be particularly important as CC analyses move toward larger datasets with substantially higher statistical precision.

\end{enumerate}

From a cosmological perspective, our result independently supports the smooth evolution of $H(z)$ predicted by $\Lambda$CDM and measured by multiple probes, including CMB, BAO, and type Ia supernova observations.  
Because CC measurements are independent of the sound horizon scale $r_d$, local distance-ladder calibrations, and assumptions about the underlying cosmological model, they provide a particularly clean probe of the expansion history.  
As such, they offer a valuable complement to other measurements currently used to investigate the Hubble tension and possible departures from standard cosmology.

The principal contribution of this work, however, is methodological.  
The Bayesian framework developed here provides a flexible and fully probabilistic approach for propagating age, metallicity, and measurement uncertainties into $H(z)$ constraints, without relying on Gaussian approximations or linearized error propagation.  
This makes the method naturally adaptable to future datasets with larger statistical power but lower spectral resolution.

Finally, the analysis framework developed here is directly applicable to future spectroscopic surveys such as Euclid \citep{Euclid_2025} and DESI \citep{DESI_2026}, as well as to narrow-band photometric surveys such as PAUS and J-PAS.
These surveys will provide vastly larger samples of passive galaxies spanning wider redshift ranges and sky areas, potentially enabling CC measurements with a precision of a few percent.  
With improved passive-galaxy selection, more accurate SPS modeling, and rigorous Bayesian treatment of the age-metallicity degeneracy, photometric and spectro-photometric CC analyses may become an increasingly powerful and independent probe of cosmic expansion history over the coming decade.

\begin{acknowledgements}
We thank the referee for their constructive feedback and valuable comments, which significantly improved the quality of this work. S.P., E.G., and P.R. acknowledge the grants PID2024-156844NB-C21 and PID2022-138896NB from MICINN/MICIU/AEI (/10.13039/501100011033), Maria de Maeztu (CEX2020-001058-M) grant, which include ERDF/FEDER funds from the European Union, and the MaX-CSIC Excellence Award MaX4-SOMMA-ICE. M.S. acknowledges support by the State Research Agency of the Spanish Ministry of Science and Innovation under the grants `Galaxy Evolution with Artificial Intelligence' (PGC2018-100852-A-I00) and `BASALT' (PID2021-126838NB-I00) and the Polish National Agency for Academic Exchange (Bekker grant BPN/BEK/2021/1/00298/DEC/1). This work was partially supported by the European Union's Horizon 2020 Research and Innovation program under the Maria Sklodowska-Curie grant agreement (No. 754510). T.M. acknowledges support through the European Space Agency (ESA) Research Fellowship in Space Science.
\end{acknowledgements}

\bibliographystyle{bibtex/aa}
\bibliography{bibchrono}

\begin{appendix}

\nolinenumbers

\section{VIPERS spectroscopic selection}
\label{Appendix:A}

\subsection{Measuring spectroscopic indices}
\label{Appendix:spec_ind}

From the discussion in Section \ref{S:3}, we know that to get a purely passive sample, it is a standard practice to use spectroscopic measurements to further refine the sample. Studies such as \citep{Borghi_2022a, Moresco_2012b} do so by combining visual inspection of emission lines with measurements of the equivalent widths of O II and O III lines and excluding EW $> 5\, \text{\AA}$ galaxies. Whereas \cite{Loubser_2025b} does only the latter, since they have a very high number of DESI galaxies, and it is not feasible to visually inspect the spectra of several hundred thousand galaxies. In contrast, in this work we only use the EW of O II combined with the Ca II H/K ratio, which was introduced as a novel passivity diagnostic \citep{Moresco_2018, Borghi_2022b}. We use these spectroscopic diagnostics to further study the sample selection and to validate our results, but in the end, we rely solely on photometric data for our final sample selection.

We have a sample of 1\,138 galaxies to start with (see Section~\ref{S:3}), and we design an automated pipeline to output the EW of O II, O III, and Ca II H \& K. For this purpose, we use the \texttt{PyLick}\footnote{The code is available at \href{https://gitlab.com/mmoresco/pylick/}{https://gitlab.com/mmoresco/pylick/}.} code developed by \cite{Borghi_2022a} and closely follow their framework. This code is capable of accurately estimating spectral indices in the UV to near-IR range and has been validated with the LEGA-C survey \citep{Borghi_2022a}. For our photometry-selected sample, we input our galaxy spectra into a \texttt{PyLick}-based pipeline, which returns the required set of indices along with their corresponding errors. The continuum windows we used for different indices are available in Table~\ref{tab:spec_indices}, where the same Ca II continuum windows were also used in \citep{Borghi_2022a}. For more information about the measurement procedure and error calculation, please see \citep{Borghi_2022a}. Note that the standard sign convention for EWs is that: (i) EWs of emission lines are above the continuum and thus are assigned a negative sign; and (ii) EWs of absorption lines are below the continuum and are assigned a positive sign. In our case, we are only measuring EWs of the O II and O III emission lines; therefore, these should be assigned negative signs. Also, if we find measurements with EW > 0, we exclude them from our sample as unreliable.

\begin{table} 
    \centering
    \caption{Definitions of spectral indices and emission-line windows.}
    \label{tab:spec_indices}
    \begin{tabular}{lccc}
        \hline\hline
        Index & Central (\AA) & Blue (\AA) & Red (\AA) \\
        \hline
        Ca\,II K   & 3925.65--3945  & 3845--3880  & 3950--3954 \\
        Ca\,II H   & 3959.40--3978  & 3950--3954  & 3983--3993 \\
        \hline
        {O\,II} ($\lambda3727$)  & 3716--3732 & 3680--3710 & 3745--3775 \\
        {O\,III} ($\lambda5007$) & 5000--5014 & 4980--4995 & 5018--5035 \\
        \hline
    \end{tabular}
    \tablefoot{ Ca II continuum windows have been taken from \cite{Fanfani_2019}.}
\end{table}

Finally, we have a dataset with the spectral measurements and their errors, and we can exclude galaxies with thresholds of EW(O II) > 5 \AA\footnote{Note that we are only considering the absolute magnitude here since all our valid EW measurements have a negative sign.} and Ca II H/K > 1.1 to select a stringent passive galaxy sample. 
However, real measurements have associated errors. Therefore, we determine the probability that a galaxy meets both selection criteria given its spectroscopic EW measurements and errors, and consider a galaxy to be passive if this probability exceeds a given threshold. The Bayesian inference of this passivity probability is described next. Using these probabilities, we can select our final sample based on the purity level we need (e.g., a sample of galaxies with more than a 90\% chance of being passive).

\subsection{Likelihood}
\label{Appendix:Likelihood}

In practice, Bayesian inference often involves evaluating a likelihood function, \(\mathcal{L}(\theta) = P(\mathrm{D}\mid \theta)\), where \(\theta\) denotes the set of model parameters.
If the errors in the data are Gaussian, the likelihood takes the form:
\begin{equation}
\mathcal{L}(\theta) = \frac{1}{(2\pi)^{n/2} \mid \mathbf{C}\mid ^{1/2}} 
e^{ -\frac{1}{2} \left(\mathbf{d} - \mathbf{m}(\theta)\right)^T \mathbf{C}^{-1} \left(\mathbf{d} - \mathbf{m}(\theta)\right)}
\label{eq:gaussian_likelihood}
\end{equation}
where:
\begin{itemize}
    \item \(\mathbf{d}\) is the data vector.
    \item \(\mathbf{m}(\theta)\) is the model prediction.
    \item \(\mathbf{C}\) is the covariance matrix of the data.
\end{itemize}

The covariance matrix \(\mathbf{C}\) encodes the uncertainties and correlations between data points:
\begin{equation}
C_{ij} = \langle (d_i - \mu_i)(d_j - \mu_j) \rangle
\label{eq:cov}
\end{equation}
 
The variance is the diagonal element: \(C_{ii} = \sigma_i^2\). In practice, sample estimates of \(\mathbf{C}\) are obtained from simulations or re-sampling methods such as jackknife or bootstrapping. The diagonal values represent sample variances if computed from data subsets.

\subsection{Selection of passive galaxies from spectroscopy}
\label{Appendix:spec_sel}

We apply Bayesian inference to the combination of our two spectroscopic indicators of passivity, Cal II H/K ($i=1$) and EW of O II ($i=2$). For each galaxy, $k$, we can define a vector $\mathbf{\Delta_k}=(\Delta_1^k,\Delta_2^k)$ with 2 components:
\begin{equation}
\label{eq:delta_k}
\Delta_i^k = \frac{d_i^k-\theta_i}{\sigma_i^k}\,, \quad i=1,2 \,,
\end{equation}
where $(d_i^k,\sigma_i^k)$ are measurements and errors estimated in spectra and $\theta_i$ are the fixed threshold values that define a passive galaxy according to our criteria: i.e., $\theta_1=1.1$ for H/K and $\theta_2=5\,\text{\AA}$ for EW of O II. This can easily be generalized to more components (e.g., including the EW of the O III line).

We can estimate the sample covariance $C_{ij}$ between these two indicators $\Delta_i$ using Eq.~\eqref{eq:cov} where $i,j=1,2$ and the mean $<..>$ and $\mu_i$ are over all $N$ galaxies in the sample:
\begin{equation}\label{eq:covariance_elements}
C_{ij} = \frac{1}{N} \sum_{k=1}^{k=N} (\Delta_i^k -\mu_i) \, (\Delta_j^k -\mu_j) \quad; \quad \mu_i = \frac{1}{N} \sum_{k=1}^{k=N} \Delta_i^k\,
\end{equation}
where $\mu_i$ and $C_{ii}$ are the sample mean and sample variance of $\Delta_i$. If the Pearson correlation coefficient $C_{12}/\sqrt{C_{11}C_{22}}$ $\le 0.1$, we can neglect the effect of the covariance in the two indicators and take $\mathbf{C}$ to be equal to the identity matrix $\mathbf{C=I}$, which will simplify the integral below. Note that $\mathbf{C}$ is the normalized covariance matrix for the $\mathbf{\Delta_k}$ defined in Eq.~\eqref{eq:delta_k}, whereas $C_{ij}$ defined in Eq.~\eqref{eq:covariance_elements} are the standard sample covariance used here only to calculate the Pearson correlation coefficient. 

The probability $P_k$ for a given galaxy $k$ to be passive is then:
\begin{equation}
P_k = \frac{\int_{\Delta_1=\Delta_1^k}^{\Delta_1=\infty}  d\Delta_1 \,\int_{\Delta_2=\Delta_2^k}^{\Delta_2=+\infty} d\Delta_2 \quad
e^{ -\frac{1}{2} \mathbf{\Delta}^T \mathbf{C}^{-1}\mathbf{\Delta} }}
{\int_{\Delta_1=-\infty}^{\Delta_1=+\infty} d\Delta_1 \, \int_{\Delta_2=-\infty}^{\Delta_2=+\infty} d\Delta_2 \quad
e^{ -\frac{1}{2} \mathbf{\Delta}^T \mathbf{C}^{-1}\mathbf{\Delta} }}
\label{eq:gaussian_likelihood2}
\end{equation}
where $\mathbf{\Delta} \equiv (\Delta_1,\Delta_2)$ are the integration variables.

When $\mathbf{C} = \mathbf{I}$, $\mathbf{C}^{-1} = \mathbf{I}$ and the exponent separates:
\begin{align}
P_k &= \frac{
\displaystyle\int_{\Delta_1 = \Delta_1^k}^{\infty} e^{-\Delta_1^2/2} d\Delta_1
\int_{\Delta_2 = \Delta_2^k}^{\infty} e^{-\Delta_2^2/2} d\Delta_2
}{
\displaystyle\int_{-\infty}^{\infty} e^{-\Delta_1^2/2} d\Delta_1
\int_{-\infty}^{\infty} e^{-\Delta_2^2/2} d\Delta_2
}
\end{align}

The passive conditions differ for each indicator:
\begin{itemize}
    \item Ca II K/H: passive if \(\mathrm{D}_1^k < \theta_1 = 1.1\)
    \item EW of O II: passive if \(\mathrm{D}_2^k < \theta_2 = 5\,\text{\AA}\)
\end{itemize}

\begin{figure}
    \centering
    \begin{subfigure}{\columnwidth}
        \includegraphics[width=\columnwidth]{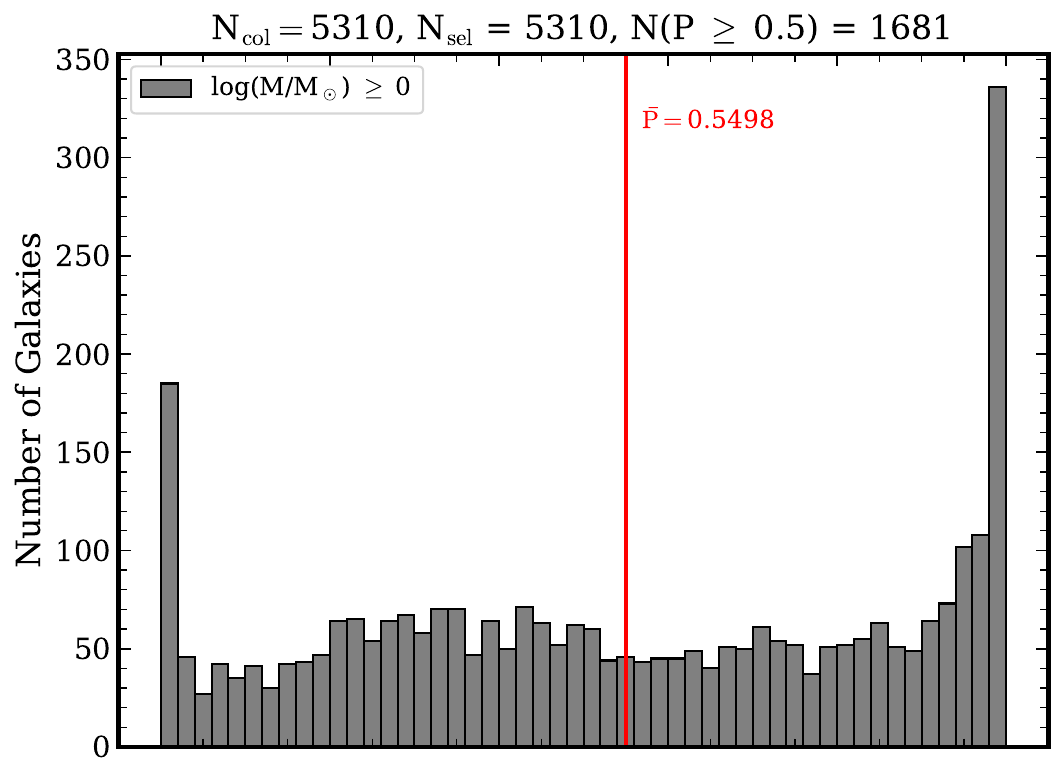}
    \end{subfigure}
    \begin{subfigure}{\columnwidth}
        \includegraphics[width=\columnwidth]{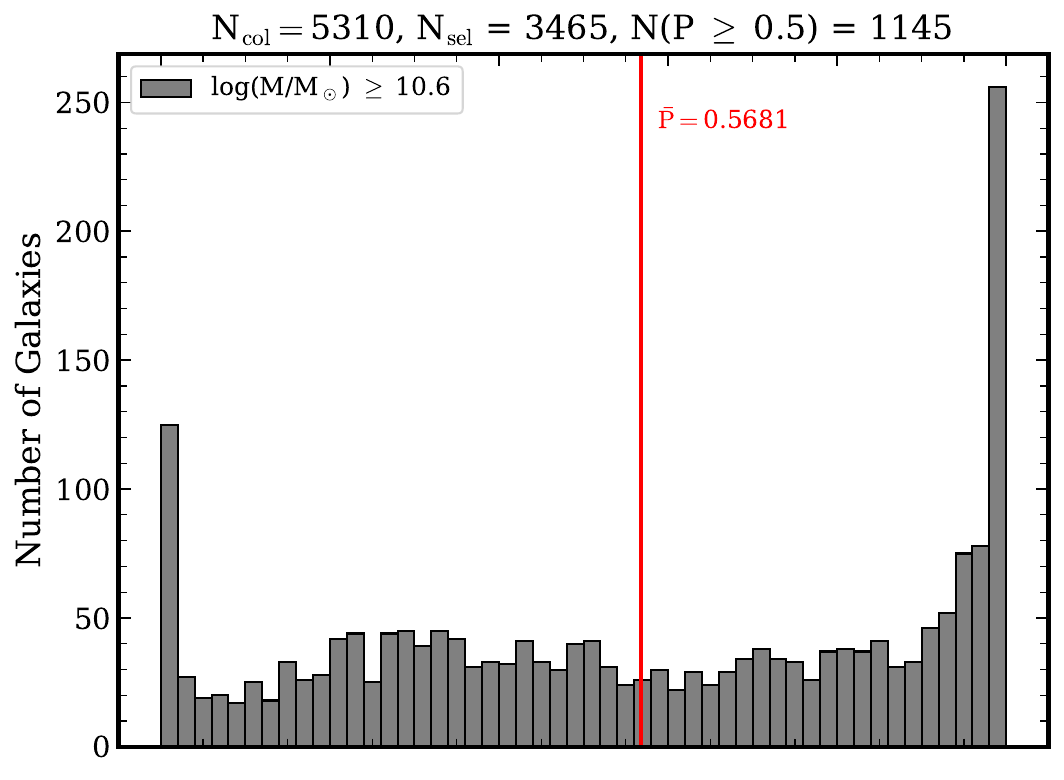}
    \end{subfigure}
    \begin{subfigure}{\columnwidth}
        \includegraphics[width=\columnwidth]{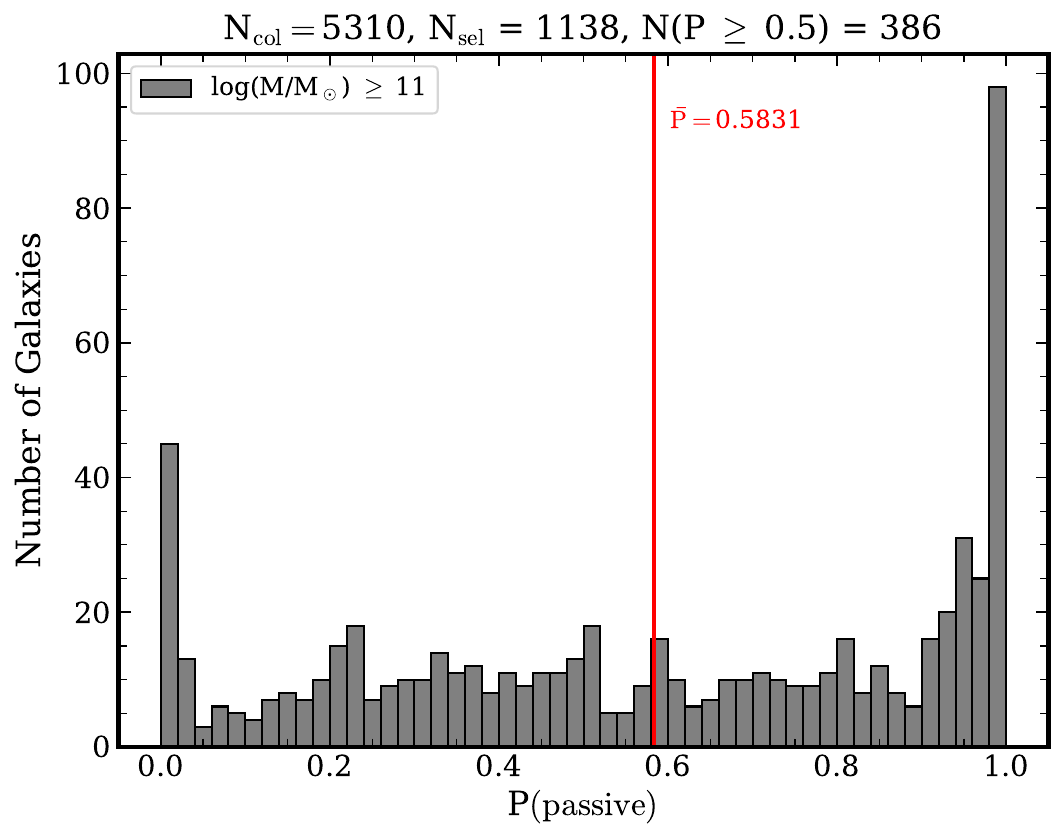}
    \end{subfigure}
    \caption{Passivity histograms for different mass cuts\textemdash $\log(M/M_\odot) \ge$ 0 (top), 10.6 (center), and 11 (bottom), with the vertical lines denoting the mean of the distribution.}
    \label{fig:P_passive_hist_mcut}
\end{figure}

\begin{figure}
    \centering
    \begin{subfigure}{0.94\columnwidth}
        \hspace{0.02\columnwidth}
        \includegraphics[width=\columnwidth]{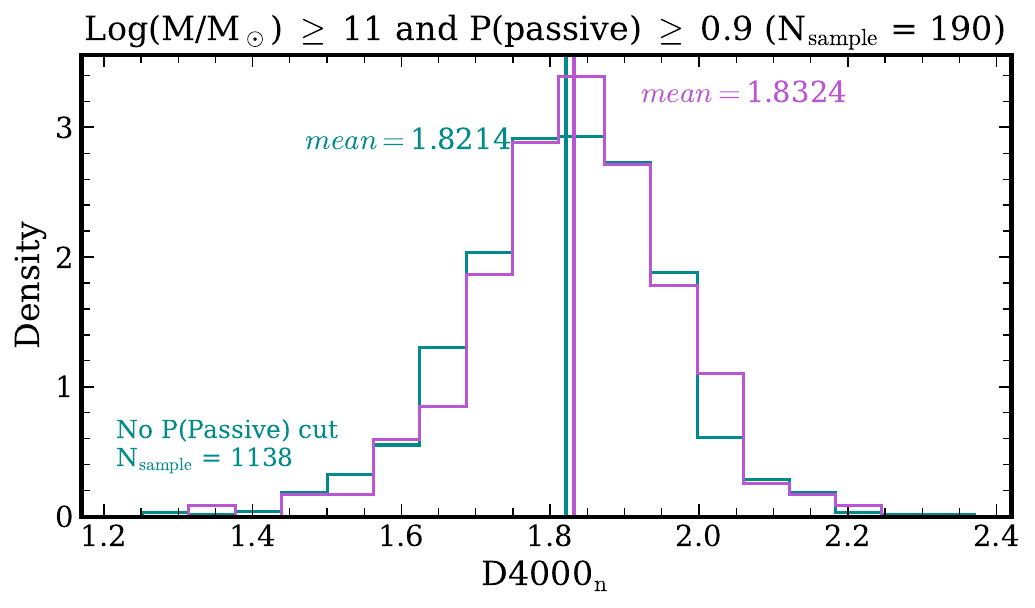}
    \end{subfigure}
    \begin{subfigure}{0.99\columnwidth}
        \includegraphics[width=\columnwidth]{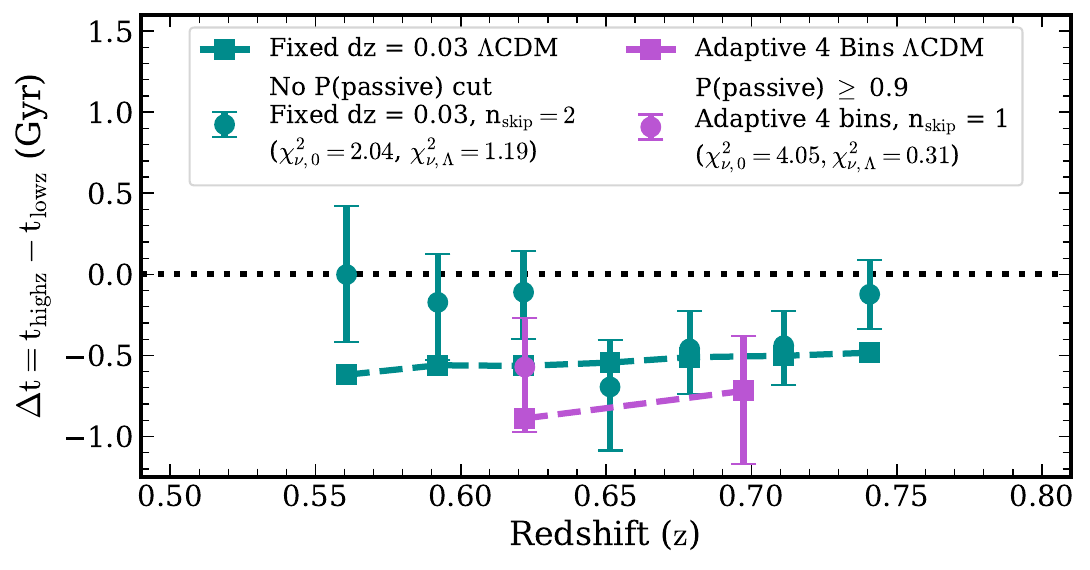}
    \end{subfigure}
    \begin{subfigure}{0.98\columnwidth}
        \hspace{0.005\columnwidth}
        \includegraphics[width=\columnwidth]{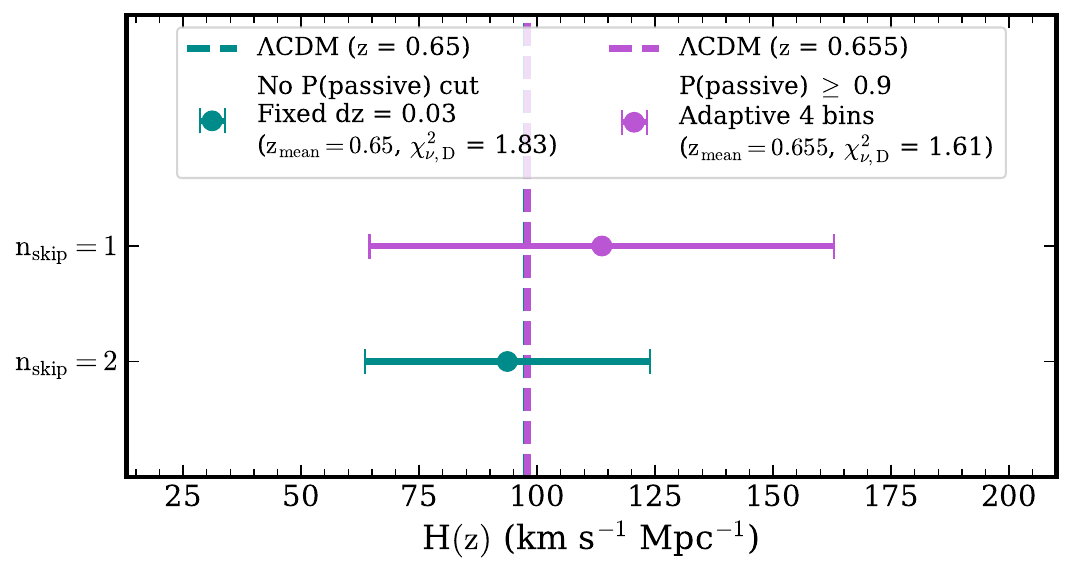}
    \end{subfigure}
    \caption{H($z$) results for a combination of a stringent mass cut and a subsequent stringent probability of passive cut from spectroscopy. The top panel features the D4000$_n$ histogram for the selected sample, the middle panel shows the time difference posterior for the best bin skip step, and the bottom panel presents the final $H(z)$ measurement for the best bin skip step.}
    \label{fig:P_passive_spec}
\end{figure}

\subsection{Measuring $H(z)$ using the spectroscopic sample}
\label{Appendix:H_spec}

Using the framework described in the previous subsection, we have a new dataset with P(passive) values for each galaxy in the color-cut-selected sample of 5310 galaxies. And in Fig.~\ref{fig:P_passive_hist_mcut}, we show how the distribution of the probability of a galaxy being passive, $P_k$ (which we also call P(passive)), looks. It is clearly a bimodal distribution, with a majority of galaxies having P(passive) > 0.9 and a considerable number having P(passive) < 0.1. As we move from the top to the bottom panels, we clearly see the effect of mass cuts, i.e., the mean probability increases from $\sim 0.55$ to $\gtrsim 0.58$. This means that both color and mass cuts improve the ``passiveness'' of our sample, while retaining a considerable amount of contaminants (for example, the $\log(M/M_\odot) \geq 11$ sample has $\sim 6\%$ galaxies with P(passive) < 0.1). 
To be as stringent as possible, we select a subsample of 190 galaxies with $\log(M/M_\odot) > 11$ and P(passive) > 0.9. We also repeat the entire $H(z)$ derivation with this very pure spectroscopic subsample, as a validation of the result.

In Fig.~\ref{fig:P_passive_spec}, we show the results from this spectroscopy-selected sample compared to the results solely from photometric cuts presented in Section~\ref{S:6}. The top panel features a $\mathrm{D}4000_n$ histogram where we see that the distributions from photometric and spectroscopic selection are slightly different in shape, but there is significant overlap. Moreover, the mean $\mathrm{D}4000_n$ value only shifts marginally from $\sim 1.82$ to $\sim 1.83$. Since we have only 190 galaxies, we divide the sample into 4 adaptive redshift bins, each with $\sim 48$ galaxies. In the middle panel of Fig.~\ref{fig:P_passive_spec}, we show the mean age differences and the corresponding asymmetric errors in comparison with our previous results from photometric selection. With $\chi^2_{\nu, 0} = 4.05$ and $\chi^2_{\nu, \Lambda} = 0.31$, we can conclusively say that our measurements are inconsistent with $\Delta t = 0$ (non-expanding universe), but the consistency with $\rm \Lambda CDM$ is still worse than the photometric one. In the bottom panel, we present the final H(z) measurement of $H(z) = 113.72 \pm 47.50 \pm 12.95$ from VIPERS spectroscopy at $z = 0.655$, which is clearly consistent with the $\rm \Lambda CDM$ prediction but has a larger error bar than the photometric counterpart due to a smaller number of objects in the sample. Moreover, with $\chi^2_{\nu,\,\mathrm{D}} = 1.61$, the scatter in the $\mathrm{D}4000_n(z)$ is low compared to the $\chi^2_{\nu,\,\mathrm{D}} = 1.83$ (which is close to 1) scatter in the measurement without any passivity cuts. However, the trade-off is the sample size and, consequently, larger error bars. 
Therefore, Fig.~\ref{fig:P_passive_spec} proves that, for the sample selection criteria we have defined in Section~\ref{S:3}, a photometric-only selection criterion provides consistent CC results, while allowing for much higher statistical power than a stringent spectroscopic selection. Moreover, the contaminants removed using the spectroscopic passivity criteria do not bias our photometric sample results.

\section{Derivation: from D4000 to Hubble parameter}
\label{Appendix:B}
\subsection{From D4000$_n$ to age posteriors: marginalization procedure}
\label{Appendix:D_to_age}

For each redshift bin with a set of passive galaxies, we infer a posterior distribution for the stellar age $t$ using the measured median D4000$_{n}$ index in that bin and its uncertainty (see Section~\ref{S:5}). We denote the observed index as $\mathrm{D}$ with Gaussian error $\sigma_{\rm D}$. The D4000$_{n}$ spectral index is modeled as a function of age $t$ and metallicity $Z$ using a precomputed stellar population synthesis grid $\mathrm{D}_{\mathrm{model}}(t_i, Z_j)$ derived in Section \ref{S:4}, which we interpolate bilinearly to obtain a continuous prediction $\tilde{\rm D}_{\mathrm{model}}(t, Z)$.

Assuming Gaussian measurement noise, the likelihood for $(t,Z)$ is
\begin{equation}
    \mathcal{L}(\mathrm{D} \mid t,Z)
    = p(\mathrm{D} \mid t,Z)
    = \frac{1}{\sqrt{2\pi}\,\sigma_{\rm D}}
      \exp\!\left[
        -\frac{\bigl(\mathrm{D} - \tilde{\rm D}_{\mathrm{model}}(t,Z)\bigr)^2}
              {2\sigma_{\rm D}^2}
      \right].
\end{equation}
We adopt independent priors for age and metallicity,
\begin{equation}
    P(t,Z) = P(t)\,P(Z),
\end{equation}
with a top-hat prior on age (in the range of 0 to 14 Gyr),
\begin{equation}
    P(t) \propto
    \begin{cases}
        1, & t_{\min} \le t \le t_{\max},\\[3pt]
        0, & \text{otherwise},
    \end{cases}
\end{equation}
and a Gaussian prior on metallicity,
\begin{equation}
    P(Z) = \mathcal{N}(Z \mid \mu_Z,\sigma_Z^2)
         = \frac{1}{\sqrt{2\pi}\,\sigma_Z}
           \exp\!\left[
             -\frac{(Z-\mu_Z)^2}{2\sigma_Z^2}
           \right],
\end{equation}
where $(\mu_Z,\sigma_Z)$ is chosen based on physically motivated external metallicity constraints (see Section~\ref{SSS:5.2.1}).

By Bayes' theorem, the joint posterior reads
\begin{equation}
    P(t,Z \mid \mathrm{D})
    \propto \mathcal{L}(\mathrm{D} \mid t,Z)\,P(t)\,P(Z).
\end{equation}
We evaluate this expression on a regular grid
$\{t_k\}_{k=1}^{N_t}$, $\{Z_\ell\}_{\ell=1}^{N_Z}$ and use
the interpolated model $\tilde{\rm D}_{\mathrm{model}}(t_k,Z_\ell)$.
For numerical stability, we compute log-likelihoods
\begin{equation}
    \ln \mathcal{L}_{k\ell}
    = -\frac{1}{2} \left(
        \frac{\mathrm{D} - \tilde{\rm D}_{\mathrm{model}}(t_k,Z_\ell)}
             {\sigma_{\rm D}}
      \right)^2
      - \ln\!\left(\sqrt{2\pi}\,\sigma_{\rm D}\right),
\end{equation}
shift them by their maximum,
\begin{equation}
    \ln \mathcal{L}'_{k\ell}
    = \ln \mathcal{L}_{k\ell}
      - \max_{k,\ell}(\ln \mathcal{L}_{k\ell}),
\end{equation}
and exponentiate to obtain a numerically well-behaved likelihood grid
$\mathcal{L}'_{k\ell} = \exp(\ln \mathcal{L}'_{k\ell})$.
Grid points outside the interpolation domain are assigned zero likelihood.

The unnormalized posterior on the grid is then
\begin{equation}
    \tilde{P}(t_k, Z_\ell \mid \mathrm{D})
    = \mathcal{L}'_{k\ell}\,P(t_k)\,P(Z_\ell).
\end{equation}
We obtain the marginal posterior in age by integrating over metallicity,
\begin{equation}\label{eq:joint_posterior}
    \tilde{P}(t_k \mid \mathrm{D})
    = \int \tilde{P}(t_k, Z \mid \mathrm{D})\,\mathrm{d}Z
    \simeq \sum_{\ell=1}^{N_Z}
           \tilde{P}(t_k, Z_\ell \mid \mathrm{D})\,\Delta Z,
\end{equation}
and normalize it so that
\begin{equation}
   P(t_k \mid \mathrm{D})
    = \frac{\tilde{P}(t_k \mid \mathrm{D})}
           {\sum_{k=1}^{N_t} \tilde{P}(t_k \mid \mathrm{D})\,\Delta t}
    \simeq \frac{\tilde{P}(t_k \mid \mathrm{D})}
                {\int \tilde{P}(t \mid \mathrm{D})\,dt}.
\end{equation}

In practice, we evaluate $P(t,Z\mid D)$ on a dense regular grid of $N_t\times N_Z=400\times400$ points spanning the age and metallicity ranges given above, using bilinear interpolation of the precomputed \texttt{BC03} D$4000_n(t,Z)$ grid (Section~\ref{S:4}) to evaluate $D_{\rm model}$ at each grid point. The marginal posterior $P(t|D)$ is obtained via trapezoidal integration over the metallicity axis (Eq.~\eqref{eq:joint_posterior}) rather than through stochastic sampling, since the low dimensionality (2D) of the parameter space and the smoothness of $D_{\rm model}$ make direct grid evaluation both exact (up to interpolation and grid-resolution error) and computationally efficient. We verified our results by comparing against MCMC posterior sampling using \texttt{emcee}.

We store this normalized posterior for use in the next steps, and from the normalized discrete posterior, we construct the cumulative
distribution function (CDF),
\begin{equation}
    F(t_k) = \sum_{i=1}^{k} P(t_i \mid \mathrm{D})\,\Delta t
    \simeq \int_{t_{\min}}^{t_k} P(t \mid \mathrm{D})\,dt,
\end{equation}
and define summary age estimates as $t_n$ which are the n$\rm ^{\rm th}$ percentile of the age distribution.
We use $t_{50}$ as the median age and the asymmetric intervals
\begin{equation}
    \Delta t_{-} = t_{\mathrm{50}} - t_{16}, \qquad
    \Delta t_{+} = t_{84} - t_{\mathrm{50}}\,,
\end{equation}
as the corresponding errors.

\subsection{From posterior ages to Hubble parameter}
\label{Appendix:age_to_H}

Given the age posteriors $P(t\mid \mathrm{D})$ in different redshift bins and the mean redshifts $z$ of the bins, we then seek the full posterior probability distribution for the age differences, $P(\Delta t\mid \mathrm{D})$, at the effective redshift $z$.

For two closely spaced bins at redshifts $z_i$, $z_{i+n}$ and ages $t_i$, $t_{i+n}$, using Eq.~\eqref{eq:cc} we can approximate with finite differences:
\begin{equation}
    H(z) \approx -\frac{1}{1+z_{\rm mid}} \frac{\Delta z}{\Delta t}\,,
\end{equation}

where $\Delta z = z_{i+n} - z_{i}$, $\Delta t = t_{i+n} - t_{i}$, $z_{\rm mid} = \frac{z_{i+n} + z_i}{2}$, $i$ is the bin index and $n$ is number of bins skipped ($n = n_{\rm skip}$).

Subsequently, from Bayesian inference, we obtain the marginalized age posteriors $P_i(t_i\mid \textrm{D}_i)$ and $P_{i+n}(t_{i+n}\mid \textrm{D}_{i+n})$ for each bin. The joint posterior is:
\begin{equation}
    P(t_i, t_{i+n} \mid \textrm{D}_i, \textrm{D}_{i+n}) = P_i(t_i \mid \textrm{D}_i)\, P_{i+n}(t_{i+n} \mid \textrm{D}_{i+n})\,.
\end{equation}

The posterior for the age difference $\Delta t = t_{i+n} - t_{i}$ is given by:
\begin{align}
    P(\Delta t \mid  D) &= \int_{-\infty}^{\infty} P_i(t_i \mid \textrm{D}_i)\, P_{i+n}(t_i + \Delta t \mid \textrm{D}_{i+n})\, dt_i\,.
\end{align}

Using this posterior distribution, we can then obtain the median $\Delta t$ value and the corresponding errors. For simplicity, we subtract the 84$^{\rm th}$ percentile and 16$^{\rm th}$ percentile asymmetric errors to get the 68$^{\rm th}$ percentile error, which can be further propagated into $H(z)$ estimates to get symmetric error bars.

Given the cosmic chronometer relation Eq.~\eqref{eq:cc}, we can plug in the $\Delta t$ and $\Delta z$ values to obtain an estimate of the Hubble parameter H$(z)$. The corresponding statistical error $\sigma_{\rm stat}$ can be calculated by propagating $\Delta t$ error, which is $\sigma_{\Delta t}$ and assuming negligible $\Delta z$ error ($\sigma_{\Delta z}\sim 0$) as:
\begin{align}
    \sigma_{\rm stat}^2 &= \left(\frac{\partial H}{\partial z}\right)^2 (\sigma_z)^2 + \left(\frac{\partial H}{\partial t}\right)^2 (\sigma_{\Delta t})^2 \\
    &= \left( \frac{1}{1+z} \frac{\Delta z}{\Delta t^2} \right)^2 (\sigma_{\Delta t})^2 \\
    &= \left(\frac{H}{\Delta t}\right)^2 \sigma_{\Delta t}^2\,.
\end{align}

\subsection{H(z) measurements from literature}\label{Appendix:H_z_table}
$H(z)$ values plotted in Fig.~\ref{fig:cc_fin} have been listed in Table \ref{tab:Hz_data}. The total error ($\sigma_{\rm H}$) is the sum of the statistical and systematic errors in quadrature ($\sigma_{\rm stat} + \sigma_{\rm syst}$). When the statistical error is large ($\sigma_{\rm stat} >> \sigma_{\rm syst}$), the total error on $H(z)$, $\sigma_{\rm H} \simeq \sigma_{\rm stat}$.

\begin{table}
\centering
\caption{$H(z)$ measurements from literature and this work.}
\label{tab:Hz_data}
\setlength{\tabcolsep}{8pt}
\renewcommand{\arraystretch}{1.25}
\resizebox{\columnwidth}{!}{
\centering
\begin{tabular}{lccc}
\hline\hline
Reference & $z$ & H$(z)$ & $\sigma_\mathrm{tot}$ \\
          &     & (km\,s$^{-1}$\,Mpc$^{-1}$) & $\sigma_\mathrm{stat} \pm \sigma_\mathrm{syst}$ \\
\hline
Moresco et al.\ (2012) & 0.179 & $75.0$   & $4.0$ \\
Moresco et al.\ (2012) & 0.199 & $75.0$   & $5.0$ \\
Moresco et al.\ (2012) & 0.352 & $83.0$   & $14.0$ \\
Moresco et al.\ (2016) & 0.380 & $81.0$   & $12.0\pm\,13.5$ \\
Moresco et al.\ (2016) & 0.400 & $85.0$   & $10.2$ \\
Moresco et al.\ (2016) & 0.425 & $87.1$   & $11.2$ \\
Moresco et al.\ (2016) & 0.445 & $92.8$   & $12.9$ \\
Moresco et al.\ (2016) & 0.478 & $80.9$   & $9.0$ \\
Moresco et al.\ (2012) & 0.593 & $104.0$  & $13.0$ \\
Moresco et al.\ (2012) & 0.680 & $92.0$   & $8.0$ \\
Moresco et al.\ (2012) & 0.781 & $105.0$  & $12.0$ \\
Moresco et al.\ (2012) & 0.875 & $125.0$  & $17.0$ \\
Moresco et al.\ (2012) & 1.037 & $154.0$  & $20.0$ \\
Moresco (2015)         & 1.363 & $160.0$  & $33.6$ \\
Moresco (2015)         & 1.965 & $186.5$  & $50.4$ \\
Loubser (2025)         & 0.460 & $88.48$  & $0.57\pm\,12.32$ \\
Loubser (2025)         & 0.670 & $119.45$ & $6.39\pm\,16.64$ \\
Loubser (2025)         & 0.830 & $108.28$ & $10.07\pm\,15.08$ \\
Loubser et al.\ (2025)\tablefootmark{a} & 0.500 & $72.1$  & $33.9\pm\,7.3$ \\
Borghi et al.\ (2022)  & 0.700 & $98.8$  & $33.6$ \\
This work (VIPERS)     & 0.650 & $93.68$ & $28.27\pm\,10.67$ \\
\hline
\end{tabular}}
\tablefoot{Uncertainties are split into statistical $\sigma_\mathrm{stat}$) and systematic ($\sigma_\mathrm{sys}$) where available. \\
\tablefoottext{a}{The Loubser et al.\ (2025) measurement was derived using BCGs only.}}
\end{table}

\section{Effect of metallicity prior}
\label{Appendix:C}

\begin{table}
    \centering
    \caption{Impact of metallicity prior variations on the $H(z)$
    measurement at $z \simeq 0.65$ ($n_{\rm skip}=2$).}
    \label{tab:Hz_metallicity_prior}
    \resizebox{\columnwidth}{!}{
    \centering
    \begin{tabular}{lcccc}
        \hline \hline
        Prior choice &
        $H(z)$ &
        $\sigma_{\rm H}$ &
        $\Delta H/H_{\rm fid}$ &
        $\Delta H/H_{\Lambda{\rm CDM}}$ \\
        &
        $({\rm km\ s^{-1}\ Mpc^{-1}})$ &
         &
        (\%) &
        (\%) \\
        \hline
        Original $(\mu,\sigma)$ &
        93.68 &
        28.27$\pm$10.67 &
        0.00 &
        -4.08 \\
        $\mu-0.05$ dex &
        69.54 &
        22.44$\pm$7.92 &
        -25.77 &
        -28.80 \\
        $\mu+0.05$ dex &
        106.36 &
        29.28$\pm$12.11 &
        +13.54 &
        +8.90 \\
        $2\sigma$ width &
        73.07 &
        36.20$\pm$8.32 &
        -22.00 &
        -25.19 \\
        \hline
    \end{tabular}}
    \tablefoot{The fractional shifts are quoted relative to our fiducial result and to the $\Lambda$CDM prediction $H_{\Lambda{\rm CDM}} = 97.67\ {\rm km\ s^{-1}\ Mpc^{-1}}$.}
\end{table}

\begin{figure}
    \centering
    \begin{subfigure}{0.98\columnwidth}
    \hspace{0.005\columnwidth}
        \includegraphics[width=\columnwidth]{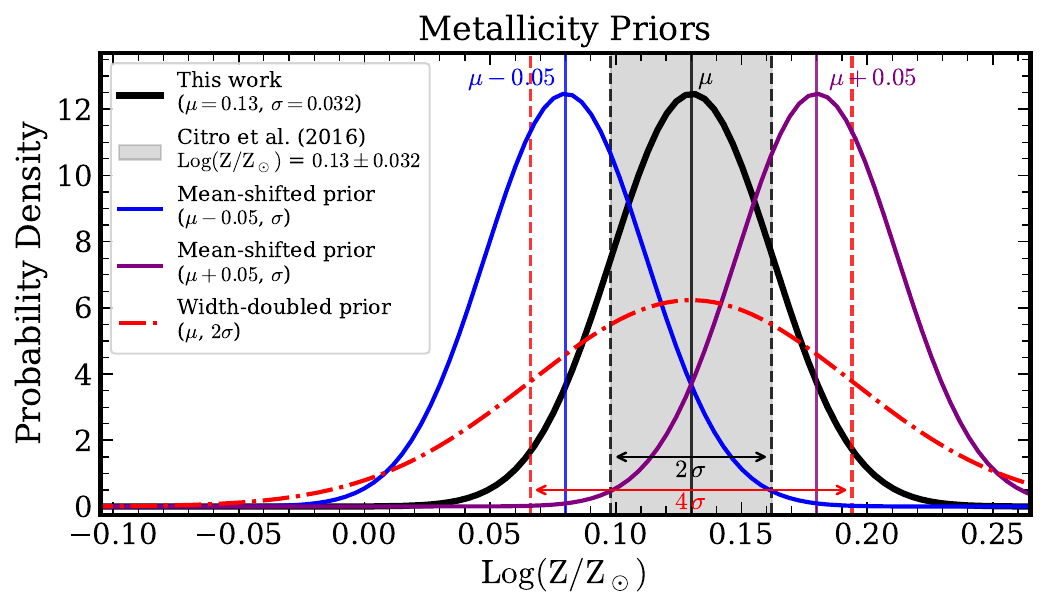}
    \end{subfigure}
    \begin{subfigure}{\columnwidth}
        \includegraphics[width=\columnwidth]{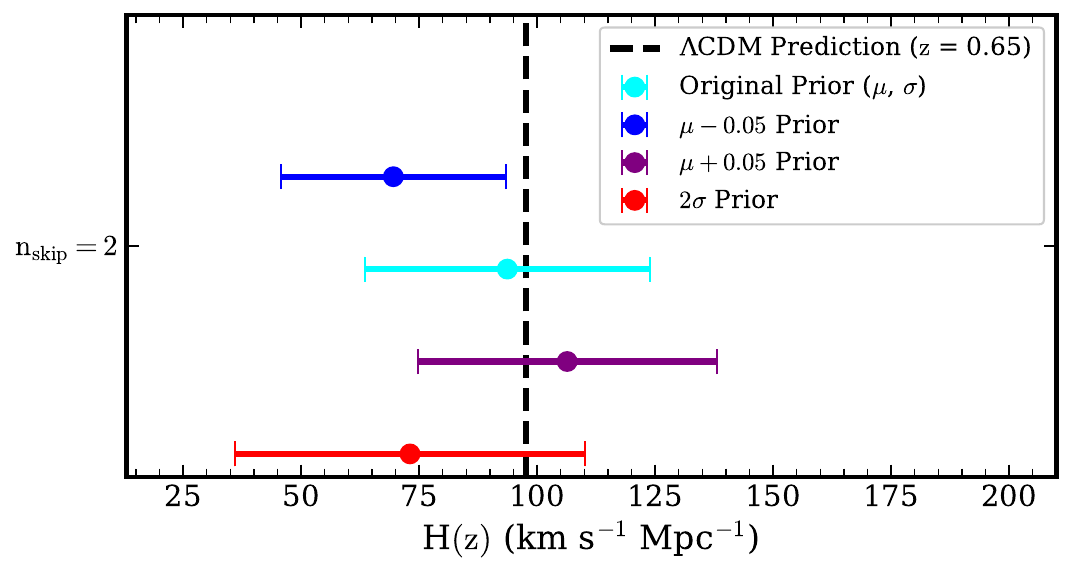}
    \end{subfigure}
    \caption{Plot of modified metallicity priors (top panel) and H(z) results for fixed $dz$ = 0.03 and $n_{\rm skip} = 2$ for the modified priors (bottom panel). $H(z)$ values come from a weighted average across the redshift range, and the corresponding error bars are a combination of statistical and systematic errors.} 
    \label{fig:prior_mod}
\end{figure}

In this section, we quantify the sensitivity of our $H(z)$ measurements to the metallicity prior used in Bayesian age inference. Our fiducial analysis adopted a Gaussian prior on the stellar metallicity, centered on the mean metallicity measured by \cite{Citro_2016} for massive and passive SDSS galaxies with a dispersion equal to their quoted measurement uncertainty (see Section~\ref{SS:5.2}). Since we are working with median D4000$_n$ values in redshift bins rather than individual galaxies, the prior represents the uncertainty in the characteristic metallicity of the CC population in each bin, and we have already fully marginalized over its effect when deriving the age posteriors.

To assess the robustness of our results to the choice of priors, we repeat the analysis for three alternative prior choices:
\begin{enumerate}
    \item a prior with the same dispersion but with the mean shifted by
    $-0.05$\,dex in $\log(Z/Z_\odot)$;
    \item a prior with the mean shifted by $+0.05$\,dex; and
    \item a prior with the same mean as the fiducial case, but with
    twice the dispersion (``$2\sigma$ width'' prior).
\end{enumerate}
These have been shown in comparison with the fiducial prior in Fig.~\ref{fig:prior_mod} (bottom panel), where we clearly see how distinct the priors are compared to the original one we chose in this work. For each of these scenarios, we recompute the age posteriors, age differences, and the corresponding constraints on $H(z)$. As an example, we focus on the binning strategy chosen for our fiducial case (fixed $dz$ = 0.03 with $n_{\rm skip} = 2$). For this case, the average or effective redshift when calculating the weighted average $H(z) = 93.68\pm28.27\pm10.67\,{\rm km\,s^{-1}\,Mpc^{-1}}$, is $z \simeq 0.65$. 

In Fig.~\ref{fig:prior_mod} (bottom panel), we show the resulting $H(z)$ measurements for the four metallicity priors (fiducial, mean shifted by $\pm0.05$\,dex, and doubled width), together with the $\Lambda$CDM prediction at the same redshift. The corresponding numerical values are listed in Table~\ref{tab:Hz_metallicity_prior}. We obtain the largest deviation from the fiducial result for the prior with the mean shifted by $-0.05$\,dex, which gives $H(z)=69.54\pm22.44\pm7.92\,{\rm km\,s^{-1}\,Mpc^{-1}}$, i.e.,\ a $\sim 26$\% decrease relative to the fiducial value and a $\sim 29$\% decrease relative to the Planck $\Lambda$CDM\ prediction at this redshift. Even in this extreme case, the measurement remains statistically consistent with both the fiducial result and $\Lambda$CDM, with differences of only $1\sigma$ and $1.2\sigma$, respectively, once the relevant uncertainties are taken into account.

The other two priors induce smaller shifts: the prior with the mean shifted by $+0.05$\,dex increases $H(z)$ by $\sim 14$\% relative to the fiducial value, while the $2\sigma$-width prior decreases it by $\sim 22$\%; in both cases, the differences are below $0.5\sigma$. We therefore conclude that, within reasonable variations around our fiducial metallicity prior, the resulting $H(z)$ constraint at $z\simeq 0.65$ is largely likelihood-dominated and only moderately sensitive to the detailed choice of metallicity prior, in agreement with previous CC robustness studies based on SPS and metallicity systematics.

\subsection{Redshift dependent metallicity prior}\label{Appendix:C.1}
\begin{figure}
    \centering
    \begin{subfigure}{\columnwidth}
        \includegraphics[width=\columnwidth]{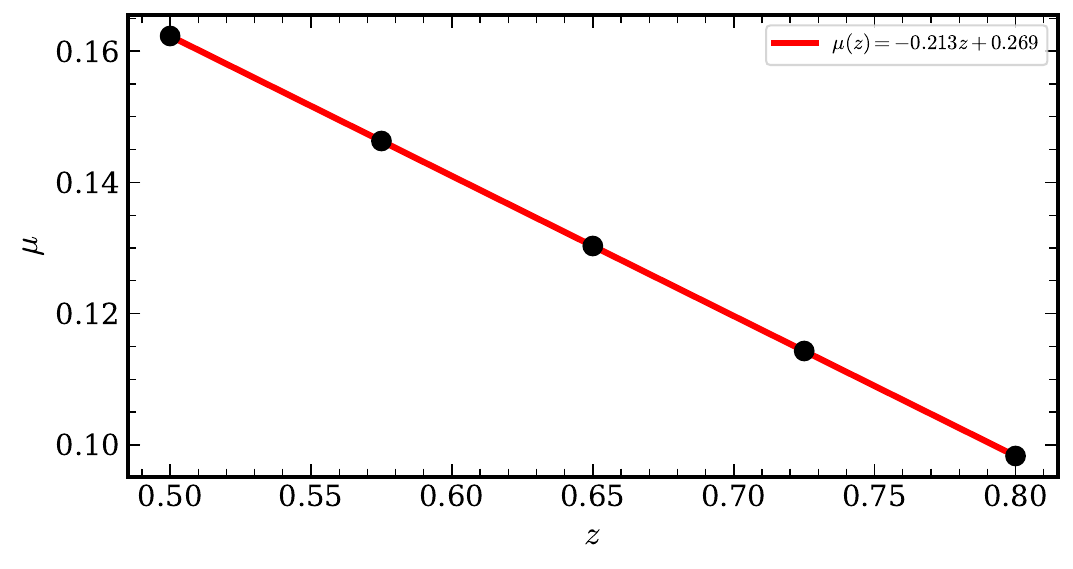}
    \end{subfigure}
    \begin{subfigure}{\columnwidth}
        \hspace{0.010\columnwidth}
        \includegraphics[width=0.98\columnwidth]{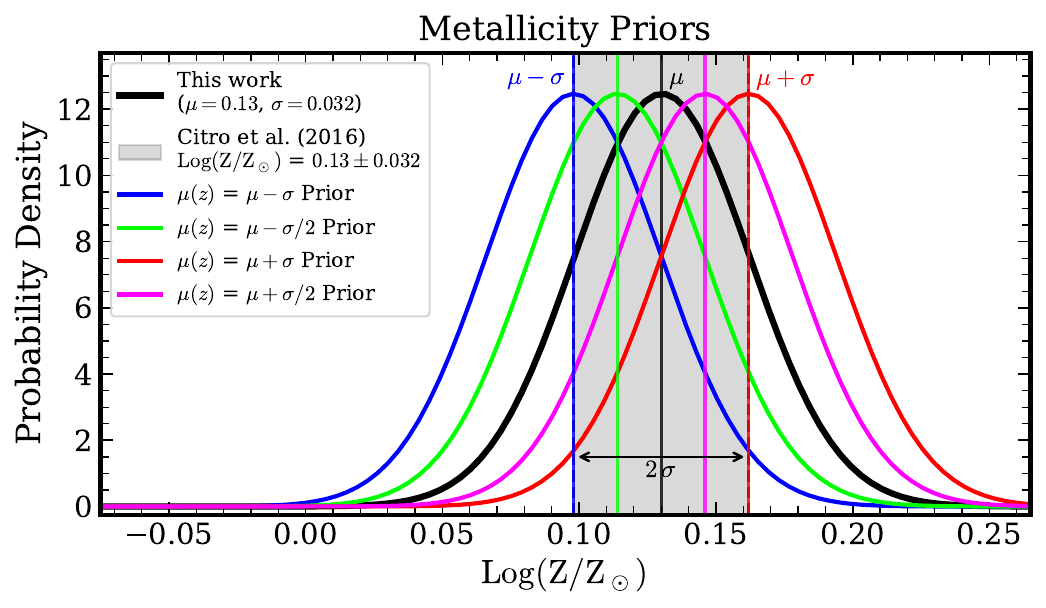}
    \end{subfigure}
    \begin{subfigure}{\columnwidth}
        \includegraphics[width=\columnwidth]{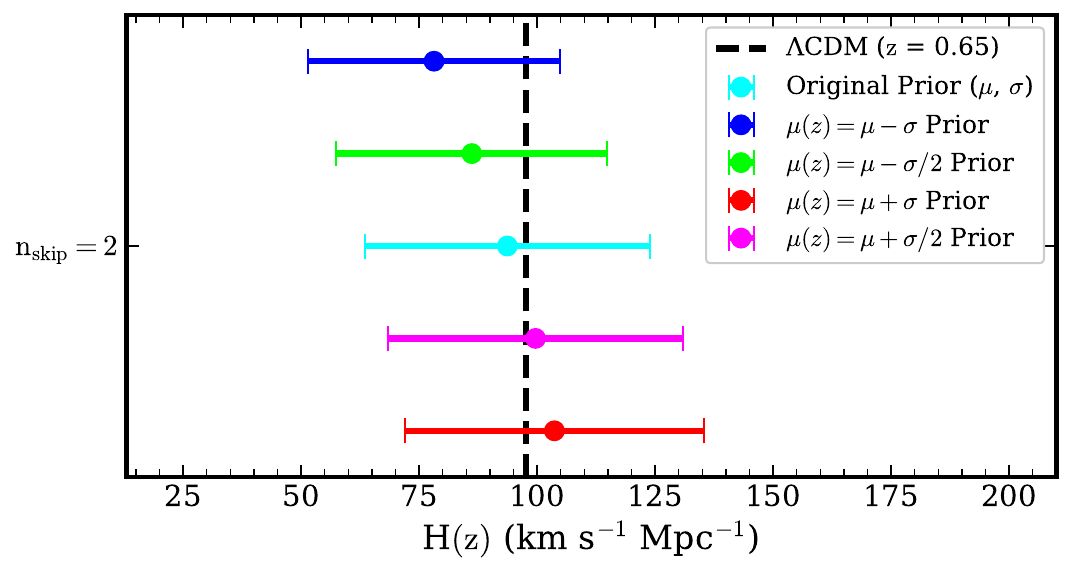}
    \end{subfigure}
    \caption{Redshift-dependent metallicity prior test. \textit{Top}: Linear trend of the metallicity prior mean with redshift, anchored to the fiducial \cite{Citro_2016} value at $z=0.65$. \textit{Middle}: Gaussian priors generated from four variants of this trend, shifted by $\pm\sigma$ and $\pm\sigma/2$ about the fiducial mean. \textit{Bottom}: Corresponding $H(z)$ measurements ($n_{\rm skip}=2$) for each variant, all consistent with the fiducial result and with $\Lambda$CDM.}
    \label{fig:red_dep_prior}
\end{figure}

In addition to the prior variations discussed above, here we test the effect of an explicit redshift-dependent metallicity prior. This is motivated by reports of mild metallicity evolution in massive passive galaxies (e.g., \citealt{Beverage_2021}), against studies finding no significant evolution (e.g., \citealt{Gallazzi_2014}, \citealt{Onodera_2015}, \citealt{Citro_2016}, \citealt{Estrada-Carpenter_2019}, \citealt{Borghi_2022a}, \citealt{Jiao_2023}). We constructed a linear trend for the prior mean (Fig.~\ref{fig:red_dep_prior}, top panel), $m_Z(z)=-0.213z+0.269$~dex, anchored to our fiducial \cite{Citro_2016} value at $z=0.65$, with slope variations spanning $\pm\sigma$ and $\pm\sigma/2$ of our fiducial dispersion (Fig.~\ref{fig:red_dep_prior}, middle panel). Recomputing $H(z)$ for each variant (Fig.~\ref{fig:red_dep_prior}, bottom panel) yields results statistically consistent with our fiducial constant-metallicity measurement and with $\Lambda$CDM, confirming that plausible metallicity evolution across $0.5\leq z\leq0.8$ does not significantly bias our inferred $H(z)$.

\section{Stellar mass evolution and the D4000$_n$ flattening at $z \sim 0.75$--$0.80$}
\label{Appendix:D}

\begin{figure}
    \centering
    \begin{subfigure}{\columnwidth}
        \includegraphics[width=\columnwidth]{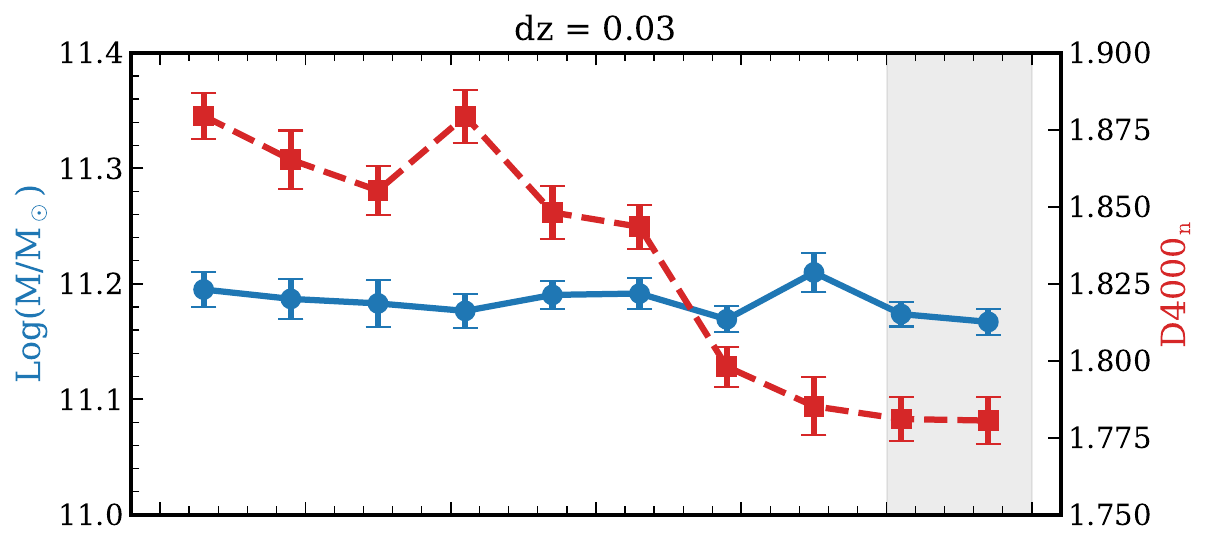}
        \label{fig:mass_d4000_dz003}
    \end{subfigure}
    \begin{subfigure}{\columnwidth}
        \includegraphics[width=\columnwidth]{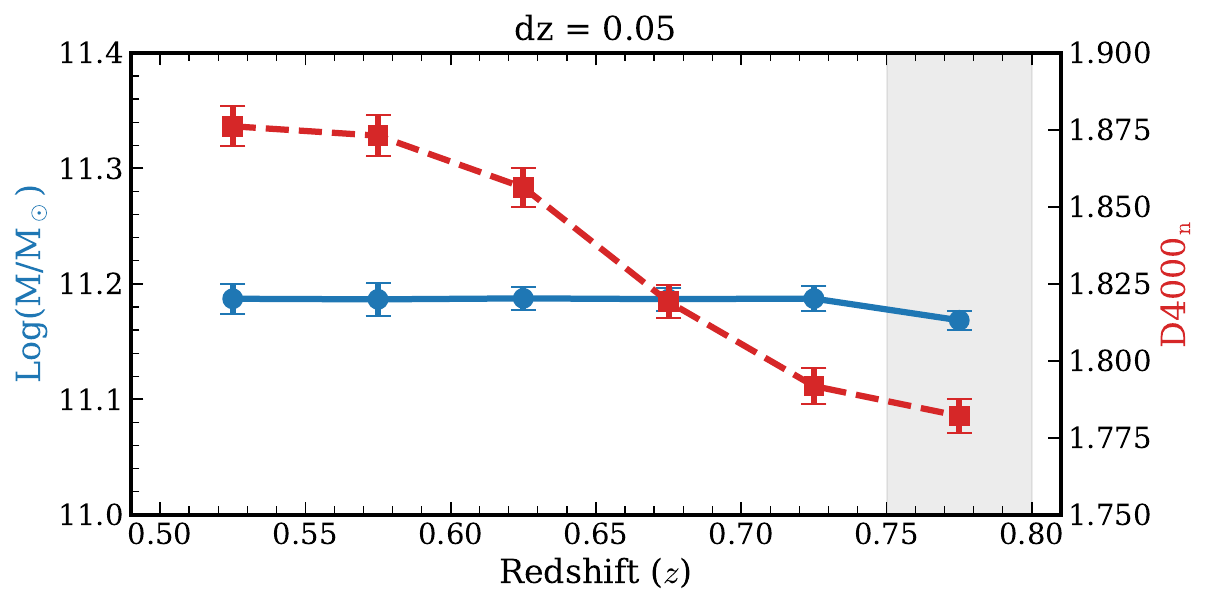}
        \label{fig:mass_d4000_dz005}
    \end{subfigure}
    \caption{Mean stellar mass and D$4000_n$ trends as a function of redshift, shown for the two bin widths adopted in this analysis: $dz = 0.03$ (top panel) and $dz = 0.05$ (bottom panel). The left (blue) and right (red) $y$-axes correspond to $\rm Log(M/M_\odot)$ and D$4000_n$, respectively.}
    \label{fig:mass_d4000_trends}
\end{figure}

To investigate whether the flattening of $\rm D4000_n$ observed near $z \approx 0.75$--$0.80$ (bottom panel of Fig.~\ref{fig:d4000_age}; Section~\ref{SSS:Age_posteriors}) is driven by a systematic variation in the stellar mass of the selected sample across redshift, we computed the mean stellar mass in the same redshift bins used for the $\rm D4000_n$ analysis. For this analysis, we adopted both bin widths considered in this work ($dz = 0.03$ and $dz = 0.05$). Figure~\ref{fig:mass_d4000_trends} shows the resulting trends.

The mean stellar mass remains nearly constant, varying only between $\approx 11.17$ and $11.22$~dex across the full redshift range $0.5 \leq z \leq 0.8$, with no systematic trend that tracks the flattening of $\rm D4000_n$ near $z \approx 0.75$--$0.80$. To quantify this, we computed the residual correlation between the binned mean stellar mass and median $\rm D4000_n$ after subtracting the overall linear trend from each quantity. We find a weak, negative correlation ($r = -0.197$ for $dz = 0.03$ and $r = -0.186$ for $dz = 0.05$). This is neither the sign nor the strength expected if a mass-driven selection effect were responsible for the observed flattening. A genuine selection bias of this kind would instead produce a positive correlation between the two residuals. Furthermore, given the small number of independent redshift bins, neither of these correlations is statistically significant ($p = 0.585$ and $p = 0.724$ for $dz = 0.03$ and $dz = 0.05$, respectively).

Therefore, we conclude that there is no evidence that the flattening of $\rm D4000_n$ near $z \approx 0.75$--$0.80$ is driven by systematic variations in stellar mass across our sample. We note that we do not have reliable velocity dispersion measurements for our photometrically selected sample, and are therefore unable to perform an equivalent check using that.

Beyond the correlation test, the near-constancy of the mean stellar mass across the full redshift range is itself indicative of a stable sample selection. If our photometric selection were increasingly biased toward more (or less) massive galaxies at higher redshift (for instance, under a mass-dependent incompleteness) we would expect a corresponding drift in mean stellar mass with $z$, which is absent. This supports the interpretation that our sample selection is approximately uniform in stellar mass across the redshift range studied, and therefore is robust to the mass downsizing scenario.

\clearpage

\end{appendix}
\end{document}